\documentclass[a4paper,twocolumn,11pt,unpublished,accepted=2024-05-05]{quantumarticle}
\pdfoutput=1
\usepackage[utf8]{inputenc}
\usepackage[english]{babel}
\usepackage[T1]{fontenc}
\usepackage{amsmath,amssymb}
\usepackage{graphicx,xcolor}% Include figure files
\usepackage{dcolumn}% Align table columns on decimal point
\usepackage{bm}% bold math
\usepackage[percent]{overpic}
\usepackage{physics}
\usepackage{mathtools}
\usepackage{hyperref}
\usepackage{booktabs}
\usepackage{ulem}

\usepackage[numbers,sort&compress]{natbib}

\newcommand\review[1]{{{ #1}}}

\begin{document}

\title{Scalable, ab initio protocol for quantum simulating SU($N$)$\times$U(1) Lattice Gauge Theories}

\author{Federica Maria Surace}
\email{fsurace@caltech.edu}
\orcid{0000-0002-1545-5230}
\affiliation{Department of Physics and Institute for Quantum Information and Matter,
California Institute of Technology, Pasadena, California 91125, USA}
\author{Pierre Fromholz}
\affiliation{The Abdus Salam International Centre for Theoretical Physics (ICTP), strada Costiera 11, 34151 Trieste,
Italy}
\affiliation{International School for Advanced Studies (SISSA), via Bonomea 265, 34136 Trieste, Italy}
\author{Francesco Scazza}
\orcid{0000-0001-5527-1068}
\affiliation{Department of Physics, University of Trieste, 34127 Trieste, Italy}
\affiliation{Istituto Nazionale di Ottica del Consiglio Nazionale delle Ricerche (CNR-INO), 34149 Basovizza-Trieste, Italy}
\author{Marcello Dalmonte}
 \affiliation{The Abdus Salam International Centre for Theoretical Physics (ICTP), strada Costiera 11, 34151 Trieste,
Italy}
\affiliation{International School for Advanced Studies (SISSA), via Bonomea 265, 34136 Trieste, Italy}

\maketitle

\begin{abstract}
  We propose a protocol for the scalable quantum simulation of SU($N$)$\times$U(1) lattice gauge theories with alkaline-earth like atoms in optical lattices. The protocol exploits the combination of naturally occurring SU($N$) pseudo-spin symmetry and strong inter-orbital interactions that is unique to such atomic species. A detailed \textit{ab initio} study of the microscopic dynamics shows how gauge invariance emerges in an accessible parameter regime, and allows us to identify the main challenges in the simulation of such theories. We provide rigorous estimates about the requirements in terms of experimental stability in relation to observing gauge invariant dynamics in both one- and two-dimensional systems, a key element for a deeper analysis on the functioning of such class of theories in both quantum simulators and computers. 
\end{abstract}

\section{Introduction}

\begin{figure}
    \centering
    \includegraphics[width=\linewidth]{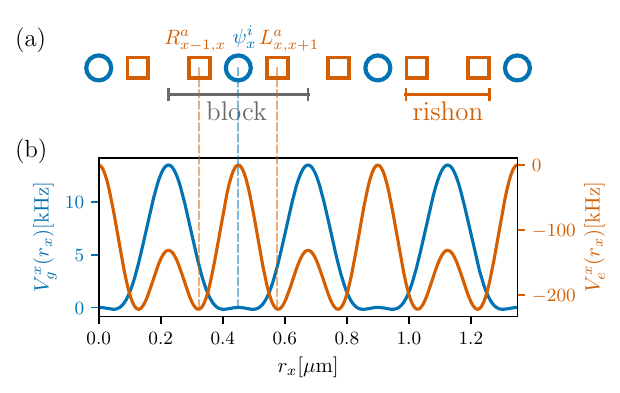}
    \includegraphics[width=0.54\linewidth]{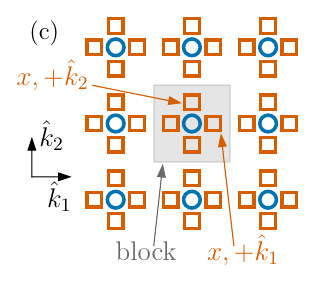}
    \includegraphics[width=0.44\linewidth]{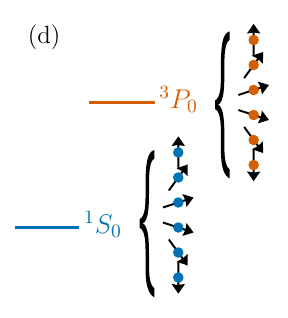}
    \caption{(a) Scheme of the 1+1d QLM: fermionic matter (in blue) resides on the sites of a one-dimensional lattice; the gauge field is representd by fermionic operators (rishons) that act on the two ends of a link (in orange). The generators of the gauge symmetry are operators acting on a single {\it block} (matter site and its neighboring gauge sites). (b) The matter/gauge sites correspond to the minima of an optical lattice for alkaline-earth-like atoms in the $g/e$ state respectively. (c) Scheme of QLM in 2+1D on a square lattice. (d) The $g/e$ states correspond to the electronic ground state (${}^1S_0$) and metastable clock state (${}^3P_0$) of the atoms. There are $N_I$ nuclear spin states decoupled from the electronic degrees of freedom.}
    \label{fig:fig1}
\end{figure}

Over the last ten years, there has been growing interest in utilizing quantum simulators to tackle the physics of gauge theories~\cite{Wiese2013,Preskill:2018aa,Zohar_2015,Dalmonte:2016jk,Banuls_2020,bauer2023quantum} - the back-bone of the standard model of particle physics. The main motivation behind this research line is the fact that gauge theories represent formidable computational challenges~\cite{Montvay1994}: while classical simulations have been tremendously successful in qualitatively and quantitatively clarifying key aspects of quantum chromodynamics (including, e.g., its phase diagram at zero chemical potential~\cite{Fukushima_2010,deTar2015} and its low-lying energy spectrum~\cite{RevModPhys.84.449}), several pivotal scenarios, such as real-time dynamics and finite baryon density, remain inaccessible to any known classical algorithm.

Starting from early theoretical proposals~\cite{Banuls_2020} and from the pioneering trapped ion experiments of Ref.~\cite{ExpPaper}, the quantum simulation of Abelian lattice gauge theories (LGTs) 
has already been shown to scale beyond simple building blocks~\cite{Anderlini_2007,Trotzky_2008,Schweizer_2019,Mil_2020,Yang_2020} 
to regimes where \review{classical simulations are challenging, with prospect of becoming prohibitive for longer timescales or larger systems}~\cite{Bernien2017,zhou2021thermalization,farrell2023scalable}. Oppositely, while digital schemes for non-Abelian LGTs have been already demonstrated in small scale experiments in Ref.~\cite{Klco_2020,Atas2021,Ciavarella2021}, no scalable non-Abelian LGTs quantum simulators exist to date. Such dearth is particularly severe in view of the motivations above, that are mostly dealing with SU$(N)$ gauge theories.  
The main roadblock for quantum simulating non-Abelian models stems from the fundamentally more complex nature of non-Abelian Gauss law in terms of quantum engineering: in particular, it is extremely challenging to constrain the dynamics of a quantum simulator in such a way that only gauge invariant states are populated. Presently, there exist only a few proposals that have addressed this \review{\cite{Banerjee2013,Zohar_2013,Zohar2013,Tagliacozzo_2013,Rico_2018,Bender_2018,Kasper2020,Kasper_2020,Mathew2022,GonzalezCuadra2022,halimeh2023spin}. In particular, Refs. ~\cite{Tagliacozzo_2013,Bender_2018,Kasper2020,Kasper_2020,GonzalezCuadra2022} employed digital or hybrid schemes, which are much more flexible but are typically expected to be harder to scale to large systems, compared to analog ones. Refs.~\cite{Rico_2018,Mathew2022} utilized explicit, exact integration of some degrees of freedom, which makes the simulation considerably simpler in one-dimension, but cannot be easily applied in higher dimensions. Finally, Refs.~\cite{Banerjee2013,Zohar_2013,Zohar2013,halimeh2023spin} are somewhat closer in spirit to our work here. They considered analogue simulation schemes for cold atoms in optical lattices, where spin-exchange or superexchange processes are used to induce the gauge-matter coupling, and gauge invariance is protected utilizing a combination of local constraints and spin symmetry. These references focused on conceptually novel aspects of realizing non-Abelian lattice gauge theory dynamics, and did not discuss how the complicated interplay of tuned lattice potentials and interactions can be mapped onto concrete physical systems. }

Here, we pursue a different approach. Building on impressive experimental developments over the last decade in harnessing quantum gases of fermionic alkaline-earth-like atoms in optical lattices \cite{Scazza14,Cappellini2014,Hofer2015,Hofrichter2016,Livi2016,Campbell2017,Riegger2018,Goban2018,Ozawa2018,ZhangZhai2020}, 
we present a proposal for the quantum simulation of SU($N$)$\times$U(1) that combines a new conceptual framework to engineer gauge invariant dynamics, with an {\it ab initio} description of the physical setup. This combined approach has two key advantages, already demonstrated along a similar route for Abelian theories~\cite{Surace2023}: firstly, it allows us to immediately identify potential challenges in terms of microscopic constraints related to the quantum hardware (e.g., inability to realize the desired background potentials, etc.), that are known to be particularly relevant in the context of quantum engineering of LGTs; and secondly, it enables us to make precise predictions in terms of energy scales of the quantum simulator, and to properly frame the theoretical issues of imperfections in realizing gauge invariant dynamics.

\section{Setup and model dynamics}

Here, we are specifically interested in realizing quantum link models (QLMs)~\cite{QLink1,QLink2,Brower1999,Chandrasekharan1997}: this formulation of lattice gauge theories is particularly advantageous for both quantum computing and simulations~\cite{wiese2022quantum} due to its finite dimensional Hilbert spaces.
 
The main ingredients of the setup we propose are schematically depicted in Fig.~\ref{fig:fig1}.
We use cold alkaline-earth fermionic atoms in their stable $^1S_0 \equiv g$ and metastable $^3P_0 \equiv e$ states, with hyperfine spin sublevels satisfying a SU($N_{\text{I}}$) symmetry~\cite{Gorshkov2010,Scazza14}, where $N_I=2I+1$ and $I$ is the nuclear spin of the atomic species. We select $N\leq N_{\text{I}}$ of these $N_{\text{I}}$ hyperfine states when initializing the system. Only these $N$ spin states are involved in the dynamics due to the global SU($N_{\text{I}}$) symmetry leading to an effective SU($N$) symmetry. The atoms are trapped via a combination of optical lattice potentials: within this setting, $g$ atoms correspond to matter fields, while double wells of $e$ atoms serve as gauge fields. In particular, the gauge field is represented by fermionic operators ({\it rishons}), and the two sites of a double well are associated with the operators acting on the left/right end of the link.

We define a \textit{block} as the ensemble of one $g$ site and its closest $e$ sites (i.e. the closest side of a double well on each adjacent link, see Fig.~\ref{fig:fig1}-(a)). With this definition, gauge invariance translates to the conservation of the number of atoms per block (the U(1) local charge) and total spin per block (the SU($N$) local charge). 
As we will detail below, the strategy to build the simulator consists in making all gauge-invariant particle configurations in each block resonant with each others, while non-gauge invariant configurations are off-resonant (or decoupled). Gauge-invariant dynamics is then realized perturbatively in this manifold, in a similar spirit as in other proposals based on linear or quadratic gauge protection \cite{Dalmonte:2016jk,halimeh2022stabilizing,Halimeh2022,halimeh2022gauge,homeier2023realistic}.

\subsection{Gauss' laws and QLM Hamiltonian}

As for any gauge theory, the key element of QLMs is the presence of local symmetries, that constrain the physical Hilbert space that shall be spanned by the engineered dynamics. Within the QLM formulation, the generators of the local symmetry can be written in terms of the left/right operators
\begin{equation}
L^a_{x,x+\hat k}=c_{x,+\hat k}^{i\dagger} \lambda_{ij}^a c_{x,+\hat k}^j
\end{equation}
\begin{equation}
R^a_{x,x+\hat k}=c_{x+\hat k,-\hat k}^{i\dagger} \lambda_{ij}^a c_{x+\hat k,-\hat k}^j.
\end{equation}

\noindent where $c_{x,+\hat k}^{i\dagger}$ is the rishon operator with color $i$ acting on the left edge of the link $x, x+\hat k$ (see Fig. \ref{fig:fig1}) and where we assume summation on the color indices. 
The matrices $\lambda_{ij}^a$, $a=1,\dots,N$, are the Pauli matrices for $N=2$ or their generalization for SU($N$) such that $\Tr \lambda^a \lambda^b = 2 \delta^{ab}$. 

We can also define the U$(1)$ electric field as
\begin{equation}
    E_{x, y} = \frac{1}{2} \left(c_{y,-\hat k}^{i\dagger}  c_{y,-\hat k}^i-c_{x,+\hat k}^{i\dagger}  c_{x,+\hat k}^i\right).
\end{equation}

With these definitions, the generators of the SU$(N)$ and U$(1)$ symmetries have the form

\begin{subequations}\label{eq:gausslaw_PRL}
\begin{align}
    G^a_x &= \psi_x^{i\dagger} \lambda_{ij}^a \psi_x^j+\sum_{\hat k} (L^a_{x, x+\hat k}+R^a_{x-\hat k, x})\nonumber\\
    &= \psi_x^{i\dagger} \lambda_{ij}^a \psi_x^j
    +\sum_{\hat{k}}\sum_{r=\pm \hat k}c_{x,r}^{i\dagger} \lambda_{ij}^a c_{x,r}^j\label{eq:SU2_gauss_law}\\
    G_x &=\psi_x^{i\dagger} \psi_x^i
    -\sum_{\hat k} (E_{x, x+\hat k}-E_{x-\hat{k}, x})-q_0 \nonumber\\
    &=\psi_x^{i\dagger} \psi_x^i
    +\sum_{\hat{k}}\sum_{r=\pm \hat k}\left(c_{x,r}^{i\dagger} c_{x,r}^i-\frac{\mathcal{N}}{2}\right)-q_0.\label{eq:U1_gauss_law}
\end{align}
\end{subequations}

The quantity $q_0$ is an additive constant that depends on the choice of the vacuum sector. 
We used $\psi_x^{i\dagger}$ (resp. $\psi_x^i$) as the creation (resp. annihilation) operator of fermionic matter with color index $i$ on site $x$.

In our implementation, they correspond to creation/annihilation operators
of a $g$ particle of SU($N$) spin $i=1,...,N$ in the $g$ well at the center of block $x$. The operators $c_{x,r}^{i\dagger}$, with $r=\pm \hat k$, are the creation operator of an $e$ particle on the $e$ well of block $x$ located on the link with block $x\pm\hat{k}$ (cf Fig.~\ref{fig:fig1}). 
All states satisfying $G_x=0$, and $G_x^a=0$ for all $a$ (called Gauss' laws) build up the gauge invariant subspace.

The target Hamiltonian of the system in the resonant subspace is the following SU($N$)$\times$U($1$) QLM Hamiltonian for massive staggered fermions (see~\cite{Wiese2013} for a review)
\begin{widetext}
\begin{align}\label{eq:allQLM}
H_{\text{QLM}}=&-\tau \sum_{\langle x, y\rangle} \left(\psi^{i \dagger}_{x}U_{x,y}^{i j} \psi^{j}_y+ \text{H.c.}\right)+ m\sum_x s_x \psi^{i \dagger}_x \psi^i_x+\frac{g^2}{2}\sum_{\langle x, y\rangle} (L_{x,y}^2+R_{x,y}^2)+\frac{{g'}^2}{2}\sum_{\langle x, y\rangle} E_{x,y}^2 \nonumber\\
&\qquad -\frac{1}{4g^2}\sum_{\Box}\Tr(U_\Box+U_\Box^\dagger)
\end{align}
\end{widetext}
with $s_x=+1$ $(-1)$ for even (odd) sites, and summation over repeated indices. The matter/anti-matter particles have mass $m$; 
The gauge assisted-hopping (matter-gauge coupling), of amplitude $\tau$, is mediated by the SU($N$) parallel transporter
$U_{x,x+\hat{k}}^{i j}= c_{x,\hat{k}}^i c_{x+\hat{k},-\hat{k}}^{\dagger j}$, according to the rishon formulation of QLM~\cite{Wiese2013}. The terms proportional to $g^2$ and $g'^2$ are the {\it electric} terms for the non-Abelian and the Abelian gauge field respectively.  We used the shorthand notation where $\langle x,y\rangle$ corresponds to the sum over the pairs of neighboring sites $x, y=x+\hat k$.  The {\it magnetic} term acts on the plaquettes of the lattice: For each plaquette spanning sites $w,x,y,z$, we defined $U_{\Box}=U_{w,x}^{ij} U_{x,y}^{jk}U_{y,z}^{kl}U_{z,w}^{li}$, where the SU$(N)$ indices $i,j,k,l$ are summed over.  We work in the representation with $\mathcal N \equiv c_{x,\hat{k}}^{i\dagger} c_{x,\hat{k}}^i+ c_{x+\hat{k},-\hat{k}}^{\dagger i}c_{x+\hat{k},-\hat{k}}^{i}=1$ rishon per link: the corresponding experimental setup has exactly one $e$ particle per double well, avoiding 2-body $e-e$ losses~\cite{Scazza14,Zhang2014}. In this representation, the electric terms are constant and have no effect on the dynamics.
We remark, however, that in our approach we will focus on implementing Gauss' law, and we will obtain an effective gauge-invariant Hamiltonian which can, in general, contain additional terms. This approach has been pursued often in quantum simulation of gauge theories~\cite{Banerjee2013}, and emphasizes the importance of gauge invariance over the exact system dynamics: it is motivated by the fact that, on the long-term goal of emulating continuum field theories, the specifics of interactions on the lattice shall play a minor role (in a similar manner as improved actions return the same continuum limit), once all local and global symmetries are faithfully realized. While we do not have continuum limit ambitions here, we still prefer to emphasize the role of symmetry over microscopic processes.

\section{Microscopic model}
We now elaborate on the derivation of a SU$(N)\times$U$(1)$ lattice gauge theory when starting from the lattice discretization of the atomic Hamiltonian. 
The continuous optical lattice Hamiltonian for alkaline-earth atoms is~\cite{Cazalilla_2009,Gorshkov2010}
\begin{widetext}
\begin{equation}\label{eq:optical_hamiltonian_PRL}
    \begin{split}
        \mathcal{H} = & \sum_{\alpha i } \int \mathrm{d}^3\mathbf{r} \Psi^\dagger_{\alpha i}\left( \mathbf{r} \right)\left(- \frac{\hbar^2}{2M}\nabla^2+V_{\alpha} \left(\mathbf{r}\right)\right)\Psi_{\alpha i}\left( \mathbf{r} \right)  + \hbar \omega_0\int \mathrm{d}^3\mathbf{r} \left(\rho_e\left(\mathbf{r}\right)-\rho_g\left(\mathbf{r}\right)\right)\\
        & \quad +\frac{g_{eg}^++g_{eg}^-}{2}\int\mathrm{d}^3\mathbf{r} \rho_e\left(\mathbf{r}\right)\rho_g\left(\mathbf{r}\right)+\sum_{\alpha,i<j}g_{\alpha \alpha}\int\mathrm{d}^3\mathbf{r} \rho_{\alpha i}\left(\mathbf{r}\right)\rho_{\alpha j}\left(\mathbf{r}\right)\\
        & \quad +\frac{g_{eg}^+-g_{eg}^-}{2}\sum_{i j}\int\mathrm{d}^3\mathbf{r} \Psi_{gi}^\dagger\left(\mathbf{r}\right)\Psi_{ej}^\dagger\left(\mathbf{r}\right)\Psi_{gj}\left(\mathbf{r}\right)\Psi_{ei}\left(\mathbf{r}\right),
    \end{split}
\end{equation}
\end{widetext}
with $\Psi^\dagger_{\alpha i}$ the field creation operator at $\mathbf{r}$ in the state $\alpha=g,e$ with SU($N$) spin $i$ and $\rho_{\alpha}\left(\mathbf{r}\right)=\sum_i\rho_{\alpha,i}\left(\mathbf{r}\right)=\sum_i\Psi^\dagger_{\alpha i}\Psi_{\alpha i}$. $M$ is the mass of the atomic species, $V_{\alpha}$ is the lattice potential, $\hbar \omega_0$ is the transition frequency between $g$ and $e$, $g_{\alpha \alpha}$, and $g_{eg}^\pm$ are the two-bodies contact interaction scattering lengths.
The lattice potential $V_{\alpha i}(\mathbf r)$ is realized using counter-propagating laser beams and is designed such that the minima of the lattice potential correspond to the lattice sites in Fig. \ref{fig:fig1}.

The lattice version of the Hamiltonian is derived by decomposing $\Psi_{\alpha i}$ into series of maximally localized Wannier functions~\cite{Kivelson1982,Marzari1997} from the populated bands of the lattices only, separated by the gap $\Delta_g$ and $\Delta_e$ from the unpopulated bands. 
Neglecting those higher bands, we are left with one Wannier function per block for the $g$ atoms and one Wannier function on each end of a link for the $e$ atoms (Figs. \ref{fig:bandsg}, \ref{fig:bandse}). The decomposition in Wannier functions of the lowest bands reads:
\begin{equation}
\label{eq:wg}
    \Psi_{gi}(\mathbf{r})=\sum_x w_g(\mathbf{r}-\mathbf{R}_x)\psi_{x}^{i},
\end{equation}

\begin{equation}
\label{eq:we}
    \Psi_{ei}(\mathbf{r})=\sum_x \sum_{s} w_{e,s}(\mathbf{r}-\mathbf{R}_{x,s})c_{x, s}^{i},
\end{equation}
where $\mathbf{R}_x$ is the center of the block $x$, and the sum over $s$ runs over all the $e$ sites in a block (for example, $s=\pm$ in 1d, $s=\pm \hat k_1, \pm \hat k_2$ in 2d). The lattice Hamiltonian is obtained by substituting Eqs. (\ref{eq:wg}) and (\ref{eq:we}) in the Hamiltonian $\mathcal H$ in Eq. (\ref{eq:optical_hamiltonian_PRL}).

\section{Gauge-invariant effective Hamiltonian}
We can write the lattice Hamiltonian as
$H=H_{0}+H_1+\dots$
where $H_0=\sum_x h_x^0$ contains all the terms acting on a single block, $H_1=\sum_{\langle xy\rangle} h^1_{x,y}$ contains all the terms connecting two neighbouring blocks, and the dots refer to all the other terms with longer range, which (as we check for the chosen realistic set of parameters in Appendix~\ref{app:abinitio}) can be neglected. In our setup, the dominant energy scale is set by the single-block Hamiltonian $H_0$, and $H_1$ can be regarded as a perturbation. This property, together with the intrinsic global SU$(N)\times$U$(1)$ symmetry of the alkaline-earth atoms, is the crucial ingredient for gauge-invariance: the global symmetries imply $[H_0, \sum_x G_x]=0$ and $[H_0, \sum_x G_x^a]=0$ for every $a$; since $h_y^0$ and the local generators $G_x$ ($G_x^a$) trivially commute for $x\neq y$, we get $[h_x^0, G_x]=0$ and $[h_x^0, G_x^a]=0$ for every $a$ and for every block $x$. In other words, $H_0$ is manifestly gauge-invariant, but is non-interacting, because it does not couple different blocks. The perturbation $H_1$, on the other hand, is not gauge invariant (it does not commute with the local generators, only with the global ones), but couples different blocks. To obtain gauge-invariant dynamics in perturbation theory, we require that all the gauge-invariant eigenstates of $H_0$ are degenerate (resonant) with respect to $H_0$: the perturbation $H_1$ can then induce non-trivial dynamics in the resonant gauge-invariant sector. Conversely, we want all gauge-variant states to be off-resonant. If some gauge-variant states are resonant, we can still obtain approximate gauge-invariant dynamics as long as the perturbation does not couple these states with the gauge-invariant sector.

\subsection{Single block Hamiltonian}

\begin{figure*}
    \centering
    \includegraphics[width=0.8\linewidth]{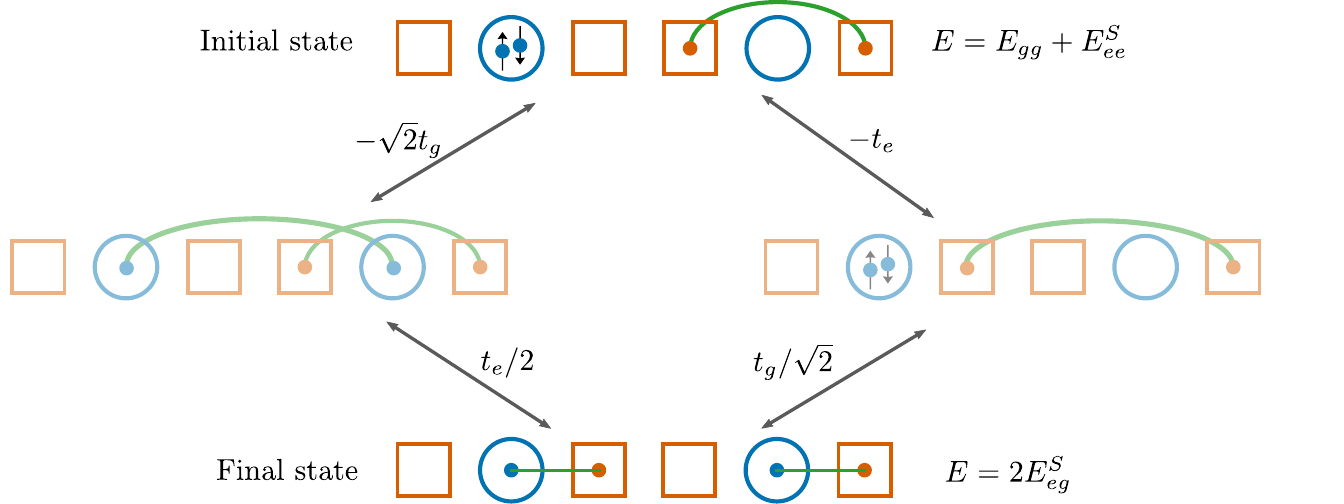}
    \caption{Example of correlated hopping obtained as a second-order process in perturbation theory. The transparent intermediate states are outside of the resonant subspace and are responsible of the second-order perturbation.}
    \label{fig:process}
\end{figure*}

\begin{table*}[t]
    \centering
    \begin{tabular}{c| c | c}
    \midrule\midrule
     \multicolumn{3}{c}{Antisymmetric nuclear spin singlet (S)}\\
        \midrule
        Graphical representation &Electronic state (sym.) & Energy  \\
        \midrule
         \raisebox{-.4\totalheight}{\includegraphics[width=2.5cm]{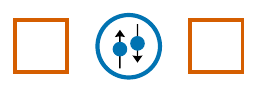}}&$\ket{g,g}$ & $E_{gg}=2\mu_g+U_{gg}$\\[1mm] 
         \raisebox{-.4\totalheight}{\includegraphics[width=2.5cm]{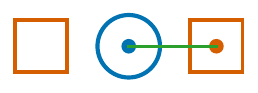}}&$\frac{1}{\sqrt{2}}\left(\ket{g,e_{r_1}}+\ket{e_{r_1}, g}\right)$ & $E_{eg}^S=\mu_g+\mu_e+U_{eg}^+$ \\[3mm]
         \raisebox{-.4\totalheight}{\includegraphics[width=2.5cm]{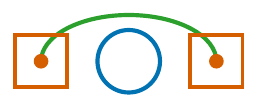}}&$\frac{1}{\sqrt{2}}\left(\ket{e_{r_1},e_{r_2}}+\ket{e_{r_2},e_{r_1}}\right)$ & $E_{ee}^S=2\mu_e$ \\[2mm]
    \midrule\midrule
    \multicolumn{3}{c}{Symmetric nuclear spin multiplet (M)}\\
    \midrule
    Graphical representation &Electronic state (antisym.) & Energy \\
    \midrule
    \raisebox{-.4\totalheight}{\includegraphics[width=2.5cm]{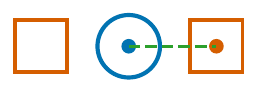}}&$\frac{1}{\sqrt{2}}\left(\ket{g,e_{r_1}}-\ket{e_{r_1}, g}\right)$ & $E_{eg}^M=\mu_g+\mu_e+U_{eg}^-$ \\[3mm]
    \raisebox{-.4\totalheight}{\includegraphics[width=2.5cm]{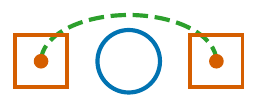}}&$\frac{1}{\sqrt{2}}\left(\ket{e_{r_1},e_{r_2}}-\ket{e_{r_2},e_{r_1}}\right)$ & $E_{ee}^M=2\mu_e$  \\[2mm]
    \midrule
    \end{tabular}
    \caption{The two particles eigenstates and eigenenergies of a single block $x$ for $N=2$. ``$e_{r_1}$'' designate a state with one $e$ particle in one of the rishons in block $x$, and $r_1,r_2$ are two different rishons. Only the singlet states satisfy gauge invariance ($G^a_x=0$ for all $a$).}
    \label{tab:states_PRL}
\end{table*}

The lattice formulation of Eq.~\ref{eq:optical_hamiltonian_PRL} involving only block $x$ is
\begin{widetext}
\begin{equation}\label{eq:hblock_PRL}
        \begin{split}
    h_x^0&=\mu_g n_x^g+\frac{U_{gg}}{2}n_x^g(n_x^g-1)+\sum_{r}\left[\mu_e n_{x,r}^e+\frac{U_{eg}^{+} +U_{eg}^{-}}{2}n_x^g n_{x,r}^e    +\frac{U_{eg}^{+} -U_{eg}^{-}}{2}\psi_x^{i\dagger} c_{x,r}^{j \dagger}\psi_x^j c_{x, r}^{i}\right]
    \\
    &\qquad +\sum_{r\neq s}t_{r s}(c_{x,r}^{i\dagger} c_{x,s}^i+\text{H.c.})+\dots,
    \end{split}
\end{equation}
\end{widetext}
with $n_x^g=\psi_x^{i\dagger}\psi_x^i$ and $n_{x,r}^e=c_{x,r}^{i\dagger}c_{x,r}^i$. The sums over $r$ and $s$ run over all the rishons within the block (i.e., the closer of the two rishons for each link connected to $x$), and the repeated spin indices $i,j$ are summed over. We assume that the $e$ double wells are separated by a large barrier, such that all hoppings $t_{rs}$ and the omitted density-assisted hoppings that change the number $\mathcal N$ of rishons per link are suppressed (see Appendix \ref{app:abinitio} for an estimate of these values for a set of realistic experimental parameters). The Hamiltonian $h_x^0$ has a U$(N)$ gauge symmetry, as shown by the commutation with the local generators $[h_x^0, G_x^a]=[h_x^0, G_x]=0$.

\subsection{Resonance conditions}
We say that a process is resonant when it connects degenerate eigenstates of $H_0=\sum_x h^0_x$. To induce the desired gauge-invariant dynamics, we require the gauge-assisted hopping to be resonant between compatible\footnote{``Compatible'' means that the configurations considered must all verify Gauss' laws simultaneously. This includes having exactly one $e$ particle on each link at all times.} gauge-invariant states.

For example for $N=2$ in one spatial dimension, the 
correlated hopping in Fig.~\ref{fig:process} is resonant if

\begin{equation}
\label{eq:eq_spectrum}
    E_{ee}^S+E_{gg}=2E_{eg}^S,
\end{equation}
with the notations of Table~\ref{tab:states_PRL}. 
This condition imposes 
$U_{gg}= 2 U_{eg}^+$. 

We now demonstrate the resonance conditions for all $N$ and all dimensions. We want to prove that all states with $G_x=0$, $G_x^a=0$ are degenerate eigenstates of $H_0=\sum_x h_x^0$, provided that $U_{gg}= 2 U_{eg}^+$. 
As shown in Appendix \ref{app:singleblock}, we can write the Hamiltonian $H_0$ as 
\begin{align}
\label{eq:H0res}
    H_0 = &\sum_x\left[ U_{eg}^+ G_x n_x-\frac{U_{eg}^+ -U_{eg}^-}{4}M^a_x G^a_x\right]\nonumber \\
    &+\frac{U_{gg}-2U_{eg}^+}{2} \sum_x  n_x (n_x-1)+\text{const.},
\end{align}
with $M^a_x= \psi^{i\dagger}_x \lambda_{ij}\psi^j_x$.
From this expression we see that, when $U_{gg}= 2 U_{eg}^+$, $H_0$ is proportional to the identity on the gauge invariant subspace, and all gauge-invariant states are degenerate.

Together with the resonance condition, we require that gauge-variant states are made off-resonant. 
We note that states with $G_x=0$, $M_x^a=0$ and $G_x^a\neq 0$ are resonant but non-gauge-invariant. An example is the gauge-breaking $e-e$ multiplet state in Table~\ref{tab:states_PRL}, which has the same energy as the $e-e$ singlet state.  
In general, when $g$-wells are allowed to be empty, such as in 1D for $N=2$, the gauge-breaking $e-e$ multiplet states have the same energy $E_{ee}^M$ as the $e-e$ gauge-invariant singlet state ($E_{ee}^S$) such that the two belong to the same degenerate subspace. However, the hybridization between the two types of states is suppressed as no on-block or nearest-block term resonantly connect the two ensembles\footnote{Because of the global SU$(N)$ symmetry, it is not possible to break Gauss' law only on a single block. If we imagine breaking Gauss' law on two neighboring blocks through a resonant second-order process, the final states of the two blocks have to be the $e-e$ multiplets. This final state has two rishon on the link between the two blocks: since the hopping between different links is negligible, these states are not connected to gauge-invariant states, which have one rishon per link. }. For $N>2$ in 1D or $N>3$ in 2D hexagonal, the available multiplet states always contain at least one $g$ particle, such that they are separated from the singlet states by an energy difference $\propto (U_{eg}^+ - U_{eg}^-)/4$. A crucial requirement for this mechanism to work is to have large energy separation $U_{eg}^+-U_{eg}^-=U_{eg}^+\cdot(1-g_{eg}^-/g_{eg}^+)$: this requirement is easily realized in fermionic alkaline-earth-like species, since the ratio of scattering lengths $g_{eg}^-/g_{eg}^+$ is very different from $1$. For example, $g_{eg}^-/g_{eg}^+=0.1170, 1.616, 0.428$ for ${}^{173}\mathrm{Yb}$, ${}^{171}\mathrm{Yb}$, and ${}^{87}\mathrm{Sr}$, respectively \cite{Scazza14,Hofer2015,Ono19,Bettermann2020,Goban2018}.

\subsection{Effective Hamiltonian}

The effective Hamiltonian can be obtained using standard perturbative methods (such as, e.g., the Schrieffer-Wolff method \cite{SW1966}). Before showing an example for the case $N=2$ in 1 spatial dimension, we are now going to discuss the general features of the effective Hamiltonian.

We assume that the dominant terms generating the dynamics in perturbation theory are single particle hoppings of $g$ particle between neighbouring blocks or $e$ particles between the two ends of a link:

\begin{align}
\label{eq:H1hop}
    H_1^\text{hop} = &-t_g \sum_{\langle x,y \rangle}(\psi^{i\dagger}_x \psi^i_{y}+\text{H.c.}) \nonumber\\
    &-t_e \sum_{x,\hat k}(c^{i\dagger}_{x,+\hat k} c^i_{x+\hat k,-\hat k}+\text{H.c.}).
\end{align}

The induced dynamics to second order in perturbation theory contains terms that act on pairs of neighboring blocks while conserving the number of rishons per link ($\mathcal N=1$) and preserving the U$(1)\times$SU$(N)$ gauge symmetry. While the effective Hamiltonian will only contain gauge-invariant terms, it will generally not coincide with the QLM Hamiltonian in Eq.~(\ref{eq:allQLM}). 

Matter-gauge coupling, for example, can be obtained at second order in perturbation theory, through a correlated hopping of the $g$ and the $e$ particle (Fig.~\ref{fig:process}). The amplitude for this process, however, depends on the configuration of the other particles in the blocks involved (see Appendix \ref{app:effH}): the corresponding Hamiltonian term is not simply given by $(\psi_x^{i\dagger} U_{x,y}^{ij} \psi_y^j +\text{H.c.})$, but will contain some dependence on the particle densities on neighboring $e$ sites. 

The mass term, which breaks space translation symmetry, can be realized by a modulation of the optical lattice potential with an additional mode, with a spatial periodicity of two blocks. The $g$ and $e$ amplitudes for this modulation are chosen to be much smaller than the ones of the main lattice potential shown in Fig.~\ref{fig:fig1}-(b), such that the Wannier functions are almost unaffected. This mode contributes to the effective lattice Hamiltonian as a site-dependent potential for the $g$ and $e$ atoms and can be chosen to be of the same order of the other terms in the effective second-order Hamiltonian. Since this mode allows to tune the mass term independently from the other parameters, we will focus, from now on, on the case $m=0$. The case $m\neq 0$ can then be straightforwardly derived, as discussed in Appendix~\ref{app:abinitio}.

Together with the matter-gauge coupling, diagonal terms (in the electric basis) are generated through second order processes. Some of these terms can be easily compensated at this perturbative order: for example an on-site interaction of the $g$ particles, of the form $n_x^2$ can be removed with a small detuning from the resonance condition $U_{gg}=2U_{eg}^+$. In order to observe some non-trivial dynamics, it is important that the diagonal terms are not too large compared to the off-diagonal correlated hopping.

Plaquette terms are the other off-diagonal terms in $H_\text{QLM}$ [Eq.~(\ref{eq:allQLM})], but they only arise through forth-order processes in perturbations theory (on a square lattice), and we therefore expect them to be very small \review{(this implies that, in 2D systems, it will be challenging to reach weak-coupling)}. Electric terms, on the other hand, are diagonal but they are trivially constant for this representation of the gauge field with one rishon per link.

As an example, we compute the effective Hamiltonian to second order in perturbation theory for the case $N=2$ in 1D:

\begin{widetext}
\begin{align}
\label{eq:heff}
    H_\text{eff}=&-\sum_x[\tau_1 +\tau_2 E_{x-1,x} E_{x+1,x+2}+\tau_3 (E_{x-1,x}- E_{x+1,x+2})](\psi_x^{i\dagger} U_{x,x+1}^{ij} \psi_{x+1}^j +\text{H.c.}) \nonumber\\
    &+\frac{u}{2}\sum_x \psi_x^{j\dagger}\psi_x^j(\psi_x^{i\dagger}\psi_x^i-1)+w\sum_{x} \psi_x^{j\dagger}\psi_x^j\psi_{x+1}^{i\dagger}\psi_{x+1}^i,
\end{align}
where

\begin{align}
    \tau_1&=\frac{ t_g t_e}{6(U^+_{eg}-U^-_{eg})}-\frac{ t_g t_e}{2(U^+_{eg}+3 U^-_{eg})}+\frac{t_g t_e}{U_{eg}^+-3U_{eg}^-},\\
    \tau_2&=\frac{ 2t_g t_e}{3(U^+_{eg}-U^-_{eg})}-\frac{ 2t_g t_e}{U^+_{eg}+3 U^-_{eg}}-\frac{4t_g t_e}{U_{eg}^+-3U_{eg}^-},
\hspace{0.9cm} \tau_3=0,\\
\label{eq:up}
    u&=\frac{8t_g^2+5t_e^2}{(U_{eg}^+-3U_{eg}^-)}-\frac{3t_e^2}{(U_{eg}^++U_{eg}^-)}+\frac{8t_g^2}{U_{eg}^+ + 3U_{eg}^-}-\frac{20t_e^2}{3(U_{eg}^+-U_{eg}^-)},\\
    w&=\frac{8t_g^2+5t_e^2}{2(U_{eg}^+-3U_{eg}^-)}-\frac{3t_e^2}{2(U_{eg}^++U_{eg}^-)}+\frac{2(t_g^2+t_e^2)}{U_{eg}^+ + 3U_{eg}^-}-\frac{4t_g^2 + 5t_e^2}{3(U_{eg}^+-U_{eg}^-)}.
\end{align}
\end{widetext}

While this effective Hamiltonian is obtained assuming that the hoppings $t_g$ and $t_e$ are the dominant terms in the perturbation, interactions between neighboring blocks can in general have a comparable amplitude: in fact, since we work in a regime with $U_{gg}, U_{eg}^+\gg t_g, t_e$, we need a strong transverse confinement, which will enhance all the interaction terms. 
A perturbative calculation that includes some of these terms is presented in Appendix \ref{app:effH}.

In general, the effective Hamiltonian obtained from perturbation theory is a useful guidance to understand the possible phases that can arise for various parameters regimes, and to easily show the interpretation of the atomic model as a non-Abelian lattice gauge theory. It stands as an open question whether the range of parameters that can be probed in the experimental setup can realize the full phase diagram \cite{SHROCK1985165,Silvi2017finitedensityphase} of the QLM in Eq.~(\ref{eq:allQLM}), and whether other phases can arise from the additional terms. Nevertheless, we expect that many of the physical properties of non-Abelian QLMs will be present in the experimental model we propose.

\section{Experimental setup}

We here discuss more in detail the experimental setup. In particular, we focus on the implementation of the optical lattice and discuss how the amplitudes of the lasers can be tuned to realize the desired parameter regime for the effective model.

\subsection{Optical lattice}

The 1D optical lattice potential displayed in Fig.~\ref{fig:fig1}-(b) can be realized by interfering laser beams at two different wavelengths $\lambda_0$, $\lambda_1$, and at angles $\theta_0, \theta_1$, creating sinusoidal potentials along the $x$-direction with amplitude $V^0_{\alpha}$ and $V^1_{\alpha}$, and periodicity $a_0\equiv\lambda_0/2\sin(\theta_0/2)$ and $a_1\equiv\lambda_1/2\sin(\theta_1/2)=a_0/2$, respectively. In addition, a tight transverse confinement can be obtained by creating an optical lattice with spacing $d$ and potential strengths $V_\alpha^{\perp}$ along the transverse $y$ and $z$ directions. Finally, a small potential $V_\alpha^S$ at angle $\theta_S$ with periodicity $a_S\equiv\lambda_S/2\sin(\theta_S/2)=2a_0$ is needed to produce the mass term. These lattices add together to give the optical potential $V_\alpha$ in Eq.~\eqref{eq:optical_hamiltonian_PRL}:
\begin{equation}
\label{eq:potential1D}
\begin{split}
        V_\alpha(\textbf{r}) = \:&V_\alpha^0 \sin^2\left(\frac{\pi x}{a_0}\right)
        +V^1_\alpha \sin^2 \left(\frac{2\pi x}{a_0}+\,\varphi \right)\\
        & +V^S_\alpha \sin^2 \left(\frac{\pi x}{ 2a_0} \right)\\
        &  
        +V_\alpha^{\perp} \left[\sin^2\left(\frac{\pi y}{d}\right)+ \sin^2\left(\frac{\pi z}{d}\right) \right].
\end{split}
\end{equation}

The wavelengths $\lambda_0, \lambda_1, \lambda_S$ of the lasers are chosen such that, for each component, the ratio between the desired amplitudes of the $g$ and $e$ potentials coincides with the ratio of the dynamical polarizabilities. We set $\theta_1=\pi$, such that the lattice spacing is $a_0=\lambda_1$. 
The phase $\varphi$ can be conveniently modulated to prepare some initial states of interest (see Appendix~\ref{sec:stateprep}), but will be otherwise set to $\varphi=0$.

The amplitudes $V_\alpha^0, V_\alpha^1, V_\alpha^S, V_\alpha^\perp$ and the width $d$ of the transverse confinement are the tunable parameters in the experimental setup (see Appendix \ref{app:optlat}). In particular, $V_\alpha^S$ determines the mass $m$ of the (staggered) fermions in the lattice gauge theory.
Note that the single-particle terms (such as the hoppings $t_g, t_e$) do not depend on the transverse confinement, while the interaction terms are enhanced when $V_\alpha^\perp$ is large: by tuning the amplitude and the width of the transverse potential with respect to the longitudinal terms $V_\alpha^0, V_\alpha^1$ we can achieve the desired regime $U_{gg}, U_{eg}^+\gg t_g, t_e$ (see Appendix~\ref{app:abinitio}). Moreover, the relative ratio between $V_g^\perp$ and $V_e^\perp$ can be adjusted to satisfy the resonance condition.

\begin{figure}[t]
    \centering
    \includegraphics[width=\linewidth]{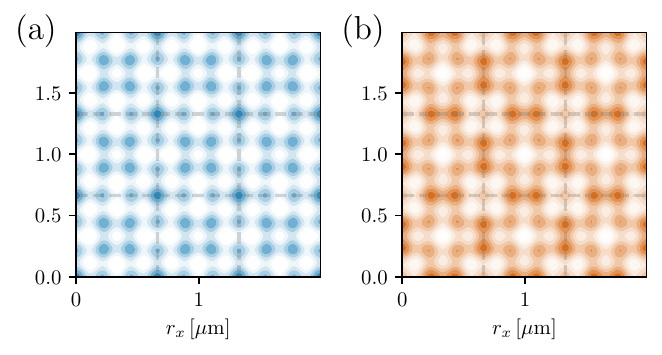}
    \includegraphics[width=0.95\linewidth]{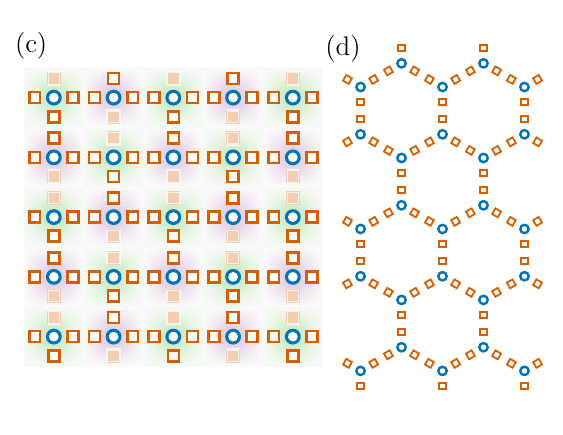}
    \caption{Optical lattice for the two-dimensional QLM: the $g$ potential (a) has minima (dark blue) in the vertices of the lattice, the $e$ potential (b) has two minima (dark orange) on each link. (c) The brickwall lattice is realized with the same potential of the square lattice ($g$ sites (blue) correspond to matter, the $e$ sites (orange) to rishons): the shaded-out orange boxes represent double wells that are initialized with no $e$ atoms, such that no gauge variable is present on the corresponding link. The staggering lattice shifts even and odd blocks (in green/purple). (d) The brickwall lattice is equivalent to the honeycomb lattice. }
    \label{fig:2D}
\end{figure}

Similar arguments can be applied in a bidimensional case.
 In Fig.~\ref{fig:2D}-(a,b) we show the profile of an optical lattice that can be used to simulate a QLM in a bidimensional square lattice:
 \begin{align}
     V_\alpha(\mathbf{r})&=V_\alpha^1 \left[ \sin^2 \left(\frac{\pi (x+y)}{a_0}\right)+\sin^2 \left(\frac{\pi (x-y)}{a_0}\right) \right]\nonumber\\
     &+V_\alpha^0 \left[ \sin^2 \left(\frac{3\pi x}{a_0}\right)+\sin^2 \left(\frac{3\pi y}{a_0}\right) \right]\nonumber\\
     &+V_\alpha^S \left[ \sin^2 \left(\frac{\pi x}{2a_0}\right)+\sin^2 \left(\frac{\pi y}{2a_0}\right) \right]\nonumber\\
     &     +V_\alpha^{\perp} \sin^2\left(\frac{\pi z}{d}\right).
 \end{align}
The superlattice potential $V_\alpha^S$ ensures the staggering of the mass term [Fig. \ref{fig:2D}-(c)]. 
Similarly to the $1D$ case, the amplitude of the transverse potential $V_\alpha^\perp$ is tuned such that the resonance condition is satisfied. 

The same optical lattice can be used to realize a brickwall lattice [equivalent to a honeycomb lattice, Fig.~\ref{fig:2D}-(d)] QLM: it is sufficient to initialize the system in a state with ``empty'' links as in Fig.~\ref{fig:2D}-(c).

\subsection{Ab initio derivation of the model}
\label{sub:abinitio}
We now show an example of realistic parameters, obtained through an ab initio calculation: the single particle Hamiltonians for the atoms in the $g/e$ optical lattices are solved using Bloch theorem; maximally localized Wannier functions are computed from the lowest band for the $g$ lattice (which is separated from the second band by an energy gap $\Delta_g$) and from the two lowest bands of the $e$ lattice (whose gap with the third band is $\Delta_e$). From the Wannier functions, we can compute the parameters in Eqs. (\ref{eq:hblock_PRL}) and (\ref{eq:H1hop}) using Eqs. \eqref{eq:optical_hamiltonian_PRL}, \eqref{eq:wg} and \eqref{eq:we}.
An example of realistic values of the experimental parameters chosen for $N=2$ in ${}^{173}$Yb, and the corresponding computed amplitudes are summarized in Tab.~\ref{tab:values1D} and considered below. The large $g_{eg}^+/g_{eg}^-$ ratio makes ${}^{173}$Yb a particularly convenient atomic species for our proposal. Note that another suitable choice is ${}^{87}$Sr, while ${}^{171}$Yb cannot be used because the negative sign of the ratio $g_{gg}/g_{eg}^+<0$ implies that the resonance condition cannot be satisfied.

\begin{table*}[t]
\centering
    \begin{minipage}[c]{0.20\linewidth}
     \begin{tabular}{p{14pt}|r}
    \midrule
     $M$ & $172.93$ u  \\ \midrule
     $g_{gg}$ & $7.748\, \mathrm{Hz}\,\mu \mathrm{m}^3$ \\ \midrule
     $g_{eg}^+$ & $72.97\, \mathrm{Hz}\, \mu \mathrm{m}^3$\\ \midrule
     $g_{eg}^-$ & $8.536 \, \mathrm{Hz}\, \mu \mathrm{m}^3$ \\ 
    \midrule
    \end{tabular} \\
    \end{minipage}
    \hfill
    \begin{minipage}[c]{0.23\linewidth}
    \begin{tabular}{p{14pt}|r}
    \midrule
     $a_0$ & $\sim 0.45 \, \mathrm{\mu m}$  \\ \midrule
     $V^0_g$ & $13.531\, h\cdot\mathrm{kHz}$  \\ \midrule
     $V^1_g$ & $-4.273\, h\cdot\mathrm{kHz}$   \\ \midrule
     $V^0_e$ & $-131.40\, h\cdot\mathrm{kHz}$   \\ \midrule
     $V^1_e$ & $-149.56\, h\cdot\mathrm{kHz}$   \\ \midrule
     $d_\perp$ & $\sim 0.80 \, \mathrm{\mu m}$  \\ \midrule
     $V_g^\perp$ &$2168.87\, h\cdot\mathrm{kHz}$    \\ \midrule
     $V_e^\perp$ & $23.29\, h\cdot\mathrm{kHz}$ \\ 
    \midrule
    \end{tabular}
    \end{minipage}
    \hfill
    \begin{minipage}[c]{0.23\linewidth}
    \flushleft
    \begin{tabular}{p{17pt}|r}
    \midrule
     $\Delta_g$ & $6.497 \, h\cdot\mathrm{kHz}$ \\ \midrule
     $\Delta_e$ & $45.821 \, h\cdot\mathrm{kHz}$\\ \midrule
     $t_g$ &  $0.040 \, h\cdot\mathrm{kHz}$\\ \midrule
     $t_e$ &  $0.118 \, h\cdot\mathrm{kHz}$\\ \midrule
     $U_{eg}^+$ & $2.565 \, h\cdot\mathrm{kHz}$  \\\midrule
     $U_{eg}^-$ & $0.300 \, h\cdot\mathrm{kHz}$  \\\midrule
     $U_{gg}$ & $5.144 \, h\cdot\mathrm{kHz}$  \\
    \midrule
    \end{tabular}
    \end{minipage}\hfill
    \begin{minipage}[c]{0.23\linewidth}
    \flushleft
    \begin{tabular}{p{10pt}|r}
    \midrule
     $u$ & $-0.014 \, h\cdot\mathrm{kHz}$ \\ \midrule
     $w$ & $ 0.012\, h\cdot\mathrm{kHz}$\\ \midrule
     $\tau_1$ &  $-0.0002\, h\cdot\mathrm{kHz}$\\ \midrule
     $\tau_2$ &  $-0.001 \, h\cdot\mathrm{kHz}$\\ \midrule
     $\tau_3$ & $0.003 \, h\cdot\mathrm{kHz}$  \\
    \midrule
    \end{tabular}
    \end{minipage}
    \caption{First column: parameters of the $^{173}$Yb in Eq.~\eqref{eq:optical_hamiltonian_PRL} \cite{Scazza14,Cappellini2014,Hofer2015}. Second column: realistic parameters for the optical lattices in Eq.~(\ref{eq:potential1D}). Third column: corresponding values of the lattice Hamiltonian,  Eqs.~\eqref{eq:hblock_PRL} and \eqref{eq:H1hop}. Fourth column: parameters of the effective Hamiltonian in Eq.~\eqref{eq:heff}.}
    \label{tab:values1D}
\end{table*}

The parameters $V_g^\perp$ and $d$ of the transverse potential are chosen to satisfy $U_{gg}\gg t_g,t_e$, such that our perturbative treatment successfully applies. We then tune $V_e^\perp$ to realize the resonance condition $U_{gg}\simeq 2U_{eg}^+$ (the interactions $U_{eg}^\pm$ grow with $V_e^\perp$, while $t_g,t_e, U_{gg}$ are unaffected). For the values in Tab.~\ref{tab:values1D}, in particular, we chose $U_{eg}^+$ such that $U_{gg}-2U_{eg}^++u=0$, i.e., such that the effective on-site matter-matter interaction vanishes up to second order in perturbation theory. We find that, since the $e$ atoms are already strongly localized in the longitudinal direction, to satisfy the resonance condition we need a stronger transverse confinement on the $g$ atoms than on the $e$ atoms. 

The desired values of the potentials $V_\alpha^1$, $V_\alpha^0$, $V_\alpha^S$ and $V_\alpha^\perp$ can be obtained by an appropriate choice of the wavelengths of the lasers, as discussed in Appendix \ref{app:optlat}.

For the values considered in Tab.~\ref{tab:values1D}, we found that the $g$ hopping mediated by $e$ particles with/without spin exchange has amplitude $(A_{eg}^- \pm A_{eg}^+)/2$ comparable to $t_g$ and $t_e$. We therefore performed a calculation of the effective Hamiltonian $H_\text{eff}$ including these terms in $H_1$. The expressions for $u,w, \tau_1, \tau_2, \tau_3$ in this case are presented in Appendix \ref{app:effH}. Their values, computed for the parameters considered here, are also reported in Tab.~\ref{tab:values1D}.

\subsection{Preparation of the initial state and readout} In order to experimentally prepare initial states within the gauge invariant subspace, where each block is occupied by a spin singlet state, one can exploit the spectroscopic resolution of the clock transition to resolve two-particle states with different electronic wavefunctions. In particular, $N=2$ gauge-invariant (spin singlet) states can be created starting from a band insulator of $g$-state atoms in the lattice potential with periodicity $a_0$, and transferring each on-site $\ket{g,g}$ (singlet) state into one of the states given in the top part of Table~\ref{tab:states_PRL}. For homogeneous or two-block translation invariant states, this can be achieved by only employing global clock laser pulses after turning on the short lattice with periodicity $a_0/2$ (see Appendix~\ref{sec:stateprep} for details). To prepare more complex gauge-invariant states such as the one depicted in Fig.~\ref{fig:dynamics}, local $g$-site addressing is required in order to manipulate site-dependent clock light shifts \cite{Weitenberg2011,Gross2021}, allowing to prepare an arbitrary singlet state starting from $\ket{g,g}$ on any given block.
Time-resolved clock spectroscopy could be exploited to measure the real-time populations of the singlet and triplet sectors, allowing for experimentally tracking the violation of SU($N$) gauge invariance. Further, U$(1)$ gauge invariance would be validated spectroscopically within the same measurement, since any single-block occupation number differing from $N$ would produce a clearly resolvable spectroscopic signal \cite{Goban2018}. The coherent evolution of each rishon-site pair, and specifically the mutually exclusive occupation of neighboring rishon sites, could be monitored using standard band-mapping techniques exploited in double-well lattices \cite{Trotzky2008}. Finally, site-resolved in situ imaging of $e$-state atoms will grant access to the electric field evolution. 

\section{Robustness of the gauge symmetries}
 In order to evaluate  the violation of gauge invariance, we 
simulated the real-time dynamics of the lattice Hamiltonian in one spatial dimension for $N=2$ using exact diagonalisation on a system of $4$ blocks with periodic boundary conditions. This method allows us  
 to check the validity of our proposal, beyond perturbative arguments.

\begin{figure}[t]
    \hfill
    \includegraphics[width=0.99\columnwidth]{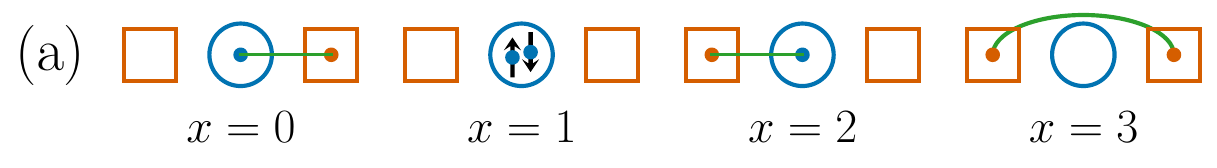}\\
    \includegraphics[width=0.99\columnwidth]{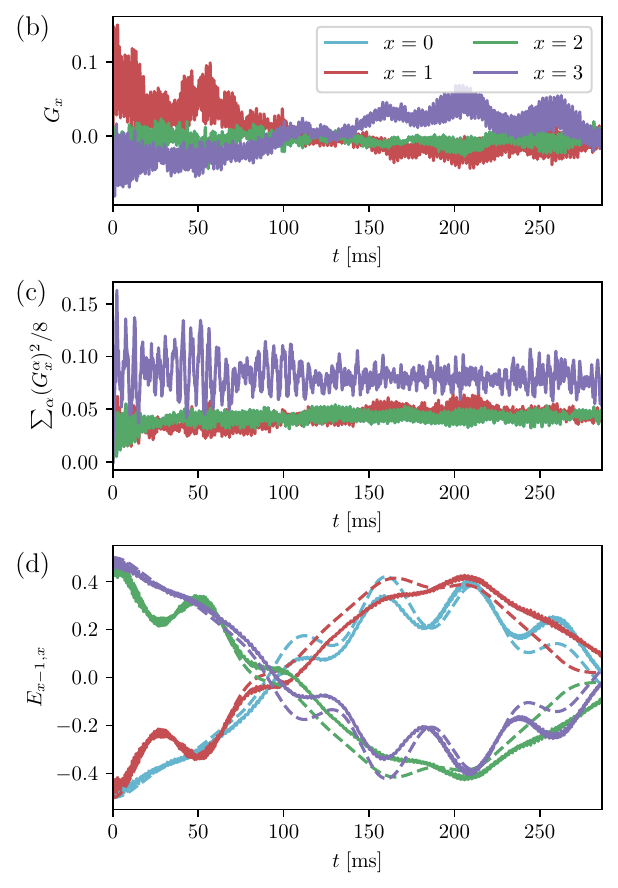}
    \caption{(a) Initial state. (b)  Abelian Gauss' law: Expectation value of the generator $G_x$. (c) Non-Abelian Gauss' law: the total gauge violation of a block $x$ is normalized such that $\sum_\alpha (G_x^\alpha)^2/8=1$ for a spin multiplet state. In (b) and (c), the lines corresponding to $x=0$ and $x=2$ perfectly overlap because of the symmetry of the initial state, so only the latter is shown.  (d) Evolution of the Abelian electric field $E_{x-1,x}$. Dashed lines: effective Hamiltonian, second-order perturbation theory.}
    \label{fig:dynamics}
\end{figure}

 In Fig. \ref{fig:dynamics}-(b-d) we plot the results for the time evolution starting from the initial state in Fig. \ref{fig:dynamics}-(a), assuming periodic boundary conditions. The simulated Hamiltonian contains $H_0$, $H_1^\text{hop}$, as well as the aforementioned density-mediated hoppings (see Appendix \ref{app:effH}). We find that both the Abelian [Fig.\ref{fig:dynamics}-(b)] and the Non-Abelian [Fig.\ref{fig:dynamics}-(c)] gauge generators show small violations of Gauss' laws, while, on the same time scale, the electric field [Fig.\ref{fig:dynamics}-(d)] has a non-trivial evolution, performing large oscillations. The evolution of the electric field is compared with the one generated by the effective Hamiltonian $H_\text{eff}$: the deviations are small at short times, and become more significant at longer times, suggesting that higher order perturbative corrections come into play (see, e.g., Ref.~\cite{Surace2023} for a similar effect). A full calculation of higher order terms is beyond the scope of our work.

While the finite system size of our simulation may have an effect on the long-time dynamics, we expect that the robustness will not depend significantly on the system size~\cite{Halimeh2020}. We note that the Gauss' law violations have a fast growth at short times (shorter than the resolution of Fig.~\ref{fig:dynamics}), but then saturate or oscillate in a bounded interval. This saturation can be understood from perturbation theory: the effective second-order Hamiltonian can be obtained from the original lattice Hamiltonian through a Schrieffer-Wolff transformation~\cite{datta1996low,BRAVYI20112793,MacDonald1988}; the same Schrieffer-Wolff transformation ``dresses'' the generators of the gauge symmetry. The ``bare'' generators differ from the dressed generators, which are quasi-conserved in the dynamics, by terms of order \review{$\lambda$}, where $\lambda$ is the perturbative parameter, leading to a saturation of the Gauss' law violations to order \review{$\lambda$}. The violations are then expected to slowly grow, departing from the saturation value at longer times, when terms beyond the second-order approximation become significant.

\section{Discussion and conclusion}
We proposed an \textit{ab initio} scalable experimental setup of an SU($N$)$\times$U(1) non-Abelian matter-gauge lattice gauge theory quantum simulator in 1D and 2D. The experiment uses ultra-cold $^{173}$Yb in an optical superlattice. The system is restricted to evolve within the resonant gauge-invariant subspace set by the initial state, while gauge-breaking states are off-resonant. For the regime of parameters we chose, the gauge-breaking dynamics is bounded and negligible. 
Our scheme thus allows for an experimental verification of gauge-invariant dynamics from first principles. 

The setup we propose can be used to investigate the real-time evolution of non-Abelian quantum link models with fermionic matter. Dynamical phenomena of interest include, for example, chiral symmetry restoration, ergodicity breaking, particle collisions, and the evolution of local defects. 

The first principle character of our investigation reveals possibilities, but also challenges along the realization of non-Abelian lattice gauge theory dynamics, that are typically not easily captured in more conceptually-minded proposals, and were thus not fully clear so far (at least, to us). In particular, similar to Abelian theories~\cite{Surace2023}, the spatial engineering of lattice potentials introduces rather complicated Wannier functions, that definitely go beyond what it is typically assumed in tight-binding approximations. Moreover, the role of exchange interactions shall be considered with great care: while those are in principle quite large energy scales, the fact that they need to be employed off-site (combined with the exotic structure of Wannier functions) makes them not as effective as expected in terms of engineering gauge invariant dynamics (this aspect was implicitly hinted to in Ref.~\cite{stannigel2014constrained}). This observation suggests that more detailed microscopic computations of effective Hamiltonians in systems involving spin exchange interactions are pivotal to quantitatively understand timescales in those promising settings~\cite{stannigel2014constrained,halimeh2023spin,Fontana2023}, as well as in those where only local interactions have been used \cite{Banerjee2013,Zohar_2013}.

At the conceptual level, an interesting perspective is to investigate alternative setups, where mixtures of alkaline-earth and alkali atoms are trapped simultaneously~\cite{singh2022dual}. It might be possible that, thanks to the much wider flexibility in trapping potentials offered by the distinct atomic structures, such systems can help circumvent some of the challenges single-species platforms face. Given recent experimental progresses in the field, investigations along this line are now, in our opinion, a promising way forward, that shall be approached with a similar first principles spirit as highlighted here.

\begin{acknowledgments}
We acknowledge useful discussions with Monika Aidelsburger.
F.~M.~S.~acknowledges support provided by the U.S.\ Department of Energy Office of Science, Office of Advanced Scientific Computing Research, (DE-SC0020290), by Amazon Web Services, AWS Quantum Program, and by the DOE QuantISED program through the theory  consortium ``Intersections of QIS and Theoretical Particle Physics'' at Fermilab. The work of M.~D. was partly supported by the MUR Programme FARE (MEPH), by QUANTERA DYNAMITE PCI2022-132919, by the PNRR MUR project PE0000023-NQSTI, and by the EU-Flagship programme Pasquans2. M.D. also thanks the Yukawa Institute for  Theoretical Physics for hospitality during the QIMG2023 workshop (YITP-T-23-01). F.~S.~acknowledges funding from the European Research Council (ERC) under the European Union’s Horizon 2020 research and innovation programme (Grant agreement No.~949438), from the Italian MUR under the FARE programme (project FastOrbit) and the PRIN programme (project CoQuS).
\end{acknowledgments}

\bibliographystyle{quantum}
\bibliography{bib}

\onecolumn
\appendix

\section{Ab-initio derivation of lattice parameters}
\label{app:abinitio}
In this Section, we will show how to explicitly derive the parameters of the lattice Hamiltonian from the amplitude of the optical potential. For a very similar discussion, see also Ref.~\cite{Surace2023}. We focus here on the case of one spatial dimension, but the procedure can also be applied to the case of two spatial dimensions, with minor changes. As explained in the main text, we can first focus on the case $m=0$ and compute the effective Hamiltonian for $V^S_\alpha=0$. The effect of $V_S^\alpha\neq 0$ will then be discussed below as a small perturbation.

\begin{figure}[h]
    \centering
\includegraphics[width=\linewidth]{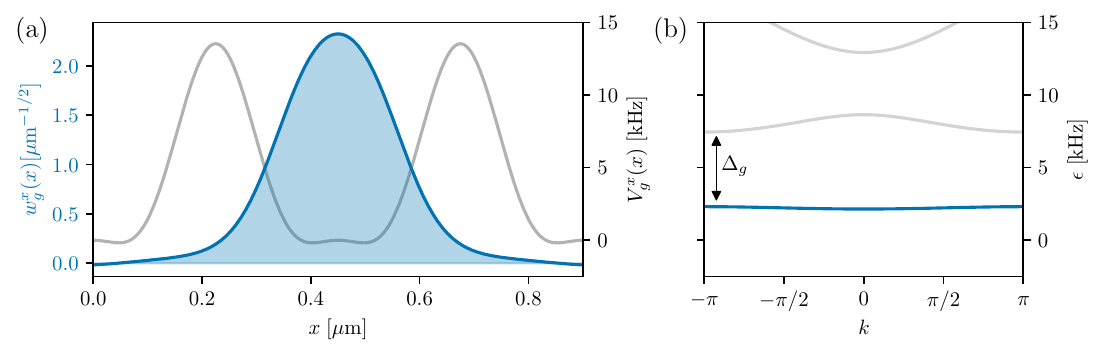}
    \caption{(a) Longitudinal lattice potential $V_g^x$ (in grey). Maximally localized Wannier function from the first Bloch band (in blue). (b) Bloch bands. }
    \label{fig:bandsg}
\end{figure}

\begin{figure}[h]
    \centering
\includegraphics[width=\linewidth]{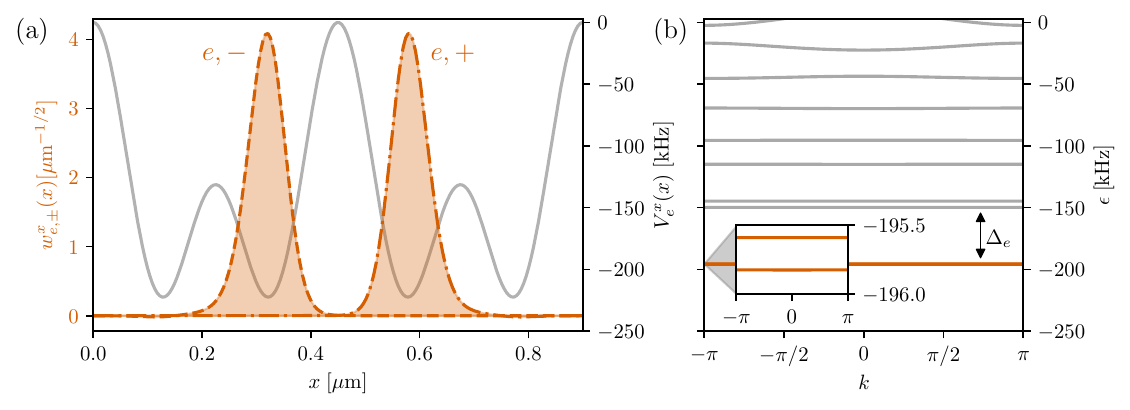}
    \caption{(a) Longitudinal lattice potential $V_e^x$ (in grey). Maximally localized Wannier functions from a combination of the first two Bloch bands (in orange). (b) Bloch bands. The inset shows the small gap between the two lowest bands. }
    \label{fig:bandse}
\end{figure}

As the first step of the procedure, we have to solve the single particle Schroedinger equation in the periodic potential of the optical lattice. Since the potential in Eq.~(\ref{eq:potential1D}) consists of a sum of $x$, $y$ and $z$ components, the solution can be factorised in the three direction. We find each of these factors separately.
We first solve the Schroedinger equation for the longitudinal potential $V_\alpha^x(x)$ for $\alpha=g,e$ [shown in grey in Figs.~\ref{fig:bandsg}-(a), \ref{fig:bandse}-(a) respectively]:
\begin{equation}
        V_\alpha^x(\textbf{r}) = V_\alpha^0 \sin^2\left(\frac{\pi x}{a_0}\right)
        +V^1_\alpha \sin^2 \left(\frac{2\pi x}{a_0}\right).
\end{equation}
The Bloch bands obtained as solutions for the $g$, $e$ atoms are shown in Figs.~\ref{fig:bandsg}-(b) and \ref{fig:bandse}-(b) respectively. 

\noindent For the $g$ lattice, the lowest band is separated from the second-lowest band by a large gap $\Delta_g$. We can focus on the lowest band to get a single localized Wannier function $w_g(x)$ per unit cell [Fig.~\ref{fig:bandsg}-(a)]. For the $e$ lattice we have to consider the two lowest bands, which are separated by a small energy gap, in order to get the two maximally localized Wannier functions $w^x_{e,-}(x)$ and $w^x_{e,+}(x)$ for the $e,-$ and $e,+$ sites in a single unit cell [Fig.~\ref{fig:bandse}-(a)]. To find the maximally localized Wannier functions we follow Refs.~\cite{Kivelson1982,Uehlinger2013} and look for the eigenstates of the position operator projected on the bands of interest.
The same procedure can be applied to compute the Wannier functions in the transverse directions. In this case, the lattice is a simple sinusoidal potential and it suffices to consider the Wannier functions $w_g^\perp$ and $w_e^\perp$ from the lowest $g$ and $e$ bands (for the values of our simulations, the potential is very deep for both species). The profile of $w_g^\perp$ and $w_e^\perp$ thus exclusively depends on the parameter $d$, $V_g^\perp$ and $V_e^\perp$

The lattice Hamiltonian can then be obtained from Eqs.~(\ref{eq:optical_hamiltonian_PRL}), (\ref{eq:wg}) and (\ref{eq:we}), with the following definitions
\begin{equation}
    w_g(x,y,z)=w_g^x(x)w_g^\perp(y)w_g^\perp(z), \qquad w_{e\pm}(x,y,z)=w_{e\pm}^x(x)w_e^\perp(y)w_e^\perp(z).
\end{equation}
We give here the definition of some of the most relevant parameters:
\begin{equation}
    t_g = -\int \mathrm{d}x\; \left(w^{x}_g(x)\right)^*\left(-\frac{\hbar^2}{2M}\partial_x^2+V_g^x(x)\right)w^x_{g}(x-a_0),
\end{equation}
\begin{equation}
    t_e = -\int \mathrm{d}x\; \left(w^{x}_{e,+}(x)\right)^*\left(-\frac{\hbar^2}{2M}\partial_x^2+V_e^x(x)\right)w^x_{e,-}(x-a_0),
\end{equation}
\begin{equation}
    U_{gg}=g_{gg}\left(\int \mathrm{d}z\; |w_g^\perp (z)|^4\right)^2 \int \mathrm{d}x\; |w_g^x (x)|^4,
\end{equation}

\begin{equation}
    U_{eg}^\pm=g_{eg}^\pm\left(\int \mathrm{d}z\; |w_g^\perp (z)|^2 |w_{e,+}^\perp (z)|^2\right)^2 \int \mathrm{d}x\; |w_g^x (x)|^2 |w_{e,+}^x (x)|^2,
\end{equation}

\begin{equation}
    A_{eg}^\pm=g_{eg}^\pm\left(\int \mathrm{d}z\; |w_g^\perp (z)|^2 |w_{e,+}^\perp (z)|^2\right)^2 \int \mathrm{d}x\; \left(w_g^x (x)\right)^*w_g^x(x-a_0) |w_{e,+}^x (x)|^2.
\end{equation}

\noindent Examples of realistic values computed using this procedure are reported in Tab.~\ref{tab:values1D}. Note however that $U_{gg}$ is of the same order of $\Delta_g$; therefore, we expect our prediction to overestimate the interaction strength, since the proximity to the higher bands effectively lowers the interactions \cite{Scazza14,Cappellini2014,Riegger2018}.

We also compute other terms, such as next-nearest neighbor hoppings, and other density-mediated terms, to show that these are negligible:
\begin{equation}
    t_g^{(1)} = -\int \mathrm{d}x\; \left(w^{x}_g(x)\right)^*\left(-\frac{\hbar^2}{2M}\partial_x^2+V_g^x(x)\right)w^x_{g}(x-2a_0),
\end{equation}
\begin{equation}
    t_e^{(1)} = -\int \mathrm{d}x\; \left(w^{x}_{e,+}(x)\right)^*\left(-\frac{\hbar^2}{2M}\partial_x^2+V_e^x(x)\right)w^x_{e,-}(x),
\end{equation}

\begin{equation}
    B_{eg}^\pm=g_{eg}^\pm\left(\int \mathrm{d}z\; |w_g^\perp (z)|^2 |w_{e,+}^\perp (z)|^2\right)^2 \int \mathrm{d}x\; \left(w_{e,+}^x (x)\right)^*w_{e,-}^x(x-a_0) |w_{g}^x (x)|^2,
\end{equation}

\begin{equation}
    C_{eg}^\pm=g_{eg}^\pm\left(\int \mathrm{d}z\; |w_g^\perp (z)|^2 |w_{e,+}^\perp (z)|^2\right)^2 \int \mathrm{d}x\; \left(w_{e,+}^x (x)\right)^*w_{e,-}^x(x-a_0) \left(w_{g}^x (x-a_0)\right)^* w_{g}^x (x),
\end{equation}

\begin{equation}
    D_{eg}^\pm=g_{eg}^\pm\left(\int \mathrm{d}z\; |w_g^\perp (z)|^2 |w_{e,+}^\perp (z)|^2\right)^2 \int \mathrm{d}x\; |w_g^x (x)|^2 |w_{e,-}^x (x-a_0)|^2,
\end{equation}

\begin{equation}
    J_{gg}=g_{gg}\left(\int \mathrm{d}z\; |w_g^\perp (z)|^4\right)^2 \int \mathrm{d}x\; |w_g^x (x)|^2|w_g^x(x-a_0)|^2,
\end{equation}
\begin{equation}
    K_{gg}=g_{gg}\left(\int \mathrm{d}z\; |w_g^\perp (z)|^4\right)^2 \int \mathrm{d}x\; \left(w_g^x (x)\right)^* \left(w_g^x (x-a_0)\right)^* w_g^x(x)^2.
\end{equation}

\noindent The values of these amplitudes for the same parameters as in Tab.~\ref{tab:values1D} are given in Tab.~\ref{tab:morevalues1D}. Note that some of these terms have non-zero matrix elements within the resonant sector of $H_0$ of our interest: $t_e^{(1)}$ for example, connects the single-block states  $\frac{1}{\sqrt{2}}(\ket{g^1 e^2_+}-\ket{g^2 e^1_+})$ and  $\frac{1}{\sqrt{2}}(\ket{g^1 e^2_-}-\ket{g^2 e^1_-})$, and would break the condition of having a single rishon ($e$ atom) per link. The error induced by this term is expected to remain small for the times $\sim 250 \; \mathrm{ms}$ considered here. The next-nearest neighbor $g$ hopping $t_g^{(1)}$, as well as the density mediated hoppings $B_{eg}^\pm$, $K_{gg}$ have no matrix elements between states in the resonant sector. Like $t_g$, $t_e$ and $A_{eg}^+$, they would contribute to the effective Hamiltonian to second order in perturbation theory, but since their amplitudes are smaller we do not expect them do have a drastic effect on the dynamics. The terms corresponding to $C_{eg}^\pm$ could directly contribute to desired the matter-gauge coupling, without invoking second-order process, but their amplitudes are so small that they have no visible effect on the time scales we consider. The terms $D_{eg}^\pm$ and $J_{gg}$ have diagonal matrix elements on gauge-invariant states and contribute as small correction to $u$ and $w$.

\begin{table*}[t]
\centering
    \begin{minipage}[c]{0.4\linewidth}
    \flushright
    \begin{tabular}{p{20pt}|r}
    \midrule
     $t_g^{(1)}$ &  $ -0.0008\, h\cdot\mathrm{kHz}$\\ \midrule
     $t_e^{(1)}$ &  $ 0.0001\, h\cdot\mathrm{kHz}$\\ \midrule
     $J_{gg}$ & $ 0.001\, h\cdot\mathrm{kHz}$  \\\midrule
     $K_{gg}$ & $ -0.023\, h\cdot\mathrm{kHz}$  \\
    \midrule
    \end{tabular}
    \end{minipage}\hfill
    \begin{minipage}[c]{0.4\linewidth}
    \flushleft
    \begin{tabular}{p{20pt}|r}
    \midrule
     $A_{eg}^+$ & $0.102 \, h\cdot\mathrm{kHz}$ \\ \midrule
     $B_{eg}^+$ & $-0.010 \, h\cdot\mathrm{kHz}$\\ \midrule
     $C_{eg}^+$ &  $-7\cdot 10^{-5}\, h\cdot\mathrm{kHz}$\\ \midrule
     $D_{eg}^+$ & $ 0.006\, h\cdot\mathrm{kHz}$  \\
    \midrule
    \end{tabular}
    \end{minipage}
    \caption{Values of additional parameters in the lattice Hamiltonian, obtained for the same values as in Tab.~\ref{tab:values1D}. For the case of $eg$ interactions, we only report the "$+$" amplitudes, as the "$-$" can be easily computed from them by multiplying by the ratio $g_{eg}^-/g_{eg}^+$.}
    \label{tab:morevalues1D}
\end{table*}

Finally, let us discuss the case $m\neq 0$, obtained with a small but non-zero $V_\alpha^S$. This term gives the following contribution to the effective lattice Hamiltonian
\begin{equation}
    H_S=\delta_g \sum_x s_x n_x^g+\delta_e \sum_x s_x (n_{x,+}^e+n_{x,-}^e),
\end{equation}
where $s_x=(-1)^x$ and
\begin{equation}
    \delta_g =\int \mathrm{d}x\; V_g^S(x) |w_g(x)|^2, \hspace{1cm} \delta_e =\int \mathrm{d}x\; V_e^S(x) |w_{e,+}(x)|^2.
\end{equation}
Using $G_x=n_x^g+\sum_r n^e_{x,r}-2$, we get
\begin{equation}
    H_S=(\delta_g-\delta_e)\sum_x s_x n_x^g+\delta_e \sum_x s_x(G_x+2).
\end{equation}
In the gauge-invariant subspace, this corresponds to a mass $m=\delta_g-\delta_e$ for the fermionic matter [see Eq.~(\ref{eq:allQLM})].

\section{Tuning of the optical lattice potential}
\label{app:optlat}
\begin{figure}
    \centering
    \includegraphics[width=0.8\linewidth]{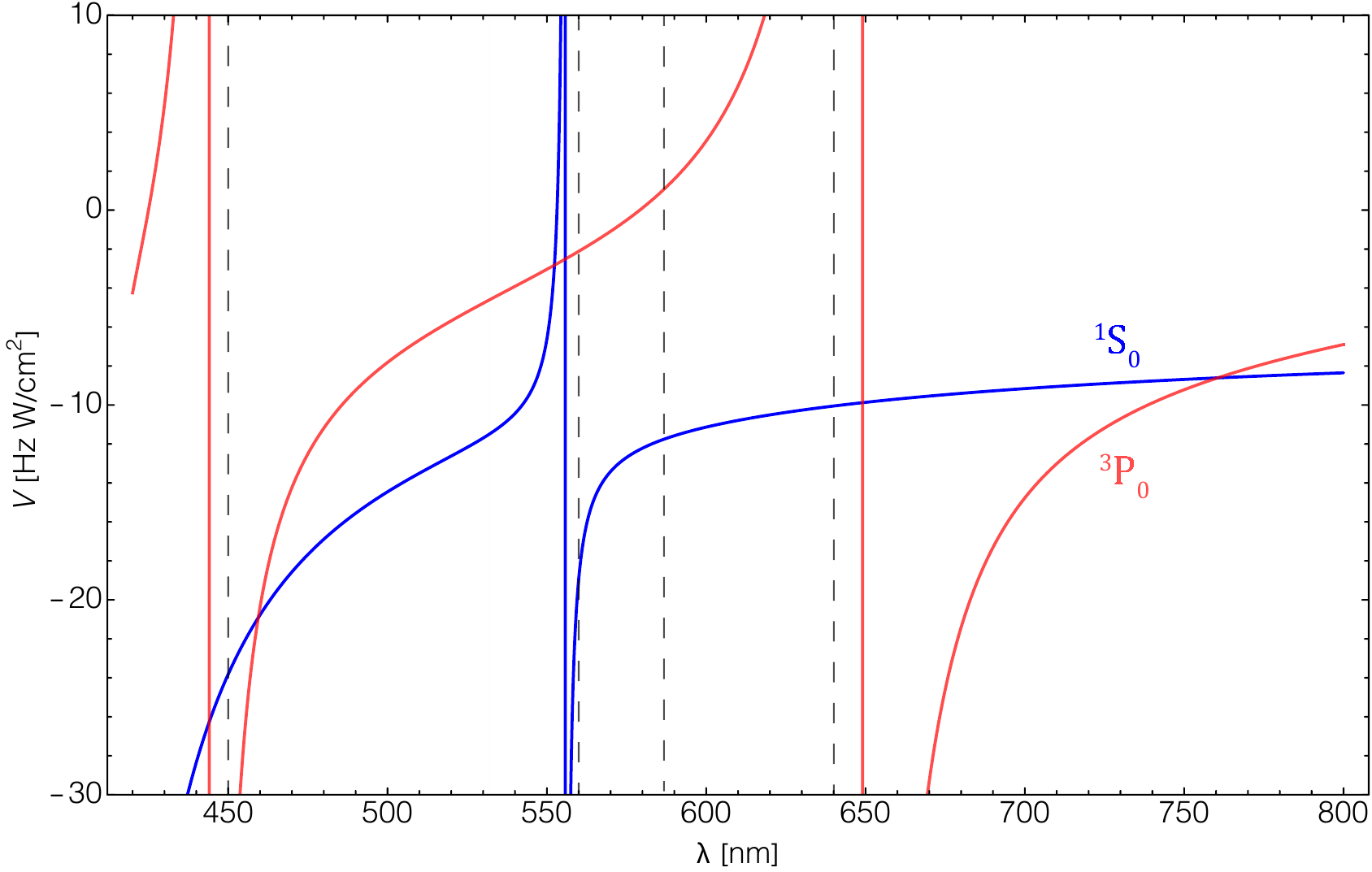}
    \caption{AC-Stark shift $V$ (normalized to the optical intensity, i.e.~in units of Hz W/cm$^2$) of the ${}^1S_0$ and ${}^3P_0$ states of ytterbium as a function of wavelength. Both curves have been calculated exploiting energy level data from the NIST Atomic Spectra database.}
    \label{fig:polarizability}
\end{figure}

The amplitudes of the terms in the lattice Hamiltonian, obtained with our ab-initio calculation, depend on six tunable parameters $V_\alpha^0$, $V_\alpha^1$, $V_\alpha^\perp$, where $\alpha=g,e$. These parameters can be independently controlled by selecting the appropriate optical intensity and the wavelength for each laser: the amplitude of the potential (for both the $g$ and $e$ atoms) is proportional to the optical intensity, while the ratio between the $g$ and $e$ amplitudes (which is determined by the ratio between the dynamical polarizability, i.e. the ratio of the state-dependent AC-Stark shifts) depends only on the wavelength. In Fig.~\ref{fig:polarizability} we plot the AC-Stark shift in units of the optical intensity for the $g\equiv{}^1 S_0$ and $e\equiv {}^3 P_0$ states of ytterbium as a function of the wavelength.

The large (positive) ratio $V_e^1/V_g^1$, required for the desired strong localization of the $e$ Wannier functions, can be achieved by choosing $\lambda_1\simeq 0.45\,\mathrm{\mu m}$ (leftmost dashed line in Fig.~\ref{fig:polarizability}), where the AC-Stark shifts have the same sign and that of the $e$ state is large, because of the proximity with a resonance. This choice, compared to other possible wavelengths that yield a similar large ratio $V_e^1/V_g^1$, has the advantage that it gives the smallest lattice spacing  $a_0=\lambda_1\simeq 0.45\, \mathrm{\mu m}$, and, consequently, the largest recoil energy. A large (in absolute value) negative $V_e^0/V_g^0$ is achieved by choosing $\lambda_0\simeq 0.64\;\mathrm{\mu m}$ (rightmost dashed line in Fig.~\ref{fig:polarizability}), while the large positive ratio of transverse confining potentials $V_g^\perp/V_e^\perp$ is obtained for $\lambda_\perp \simeq 0.56 \; \mathrm{\mu m}$ or $\lambda_\perp \simeq 0.59 \; \mathrm{\mu m}$.
The condition $a_0=\lambda_1$ can then be met by using an angle $\theta_0\simeq 0.518\cdot \pi$.

\section{Single block Hamiltonian}
\label{app:singleblock}
We now demonstrate how to derive Eq. (\ref{eq:H0res})  from the single block Hamiltonian in Eq. (\ref{eq:hblock_PRL}). We first note that, by using the identity $\lambda^a_{ij}\lambda^a_{kl}=2\delta_{il}\delta_{jk}-\frac{2}{N}\delta_{ij}\delta_{kl}$, we can write

\begin{align}
\begin{split}
    \psi_x^{i\dagger}c_{x,r}^{j^\dagger}\psi_x^j c_{x,r}^i&=-\frac{1}{2}\psi_x^{i\dagger}\lambda_{ij}\psi_x^j c_{x,r}^{k\dagger}\lambda_{kl}^a c_{x,r}^l-\frac{1}{N}\psi_x^{i^\dagger}\psi_x^i c_{x,r}^{j\dagger}c_{x,r}^j\\
    & = -\frac{1}{2}M_{x}^a(G_x^a-M_x^a)-\frac{1}{N}n_x^g n_{x,r}^e,
\end{split}
\end{align}
where $M_{x}^a=\psi_x^{i\dagger}\lambda_{ij}\psi_x^j$. The quadratic Casimir operator in $M_x^a$ can be expressed as

\begin{align}
\begin{split}
    \frac{1}{2}\sum_a (M_x^a)^2&=\psi_x^{i\dagger}\psi_x^j\psi_x^{j\dagger}\psi_x^i-\frac{1}{N} \psi_x^{i\dagger}\psi_x^i\psi_x^{j\dagger}\psi_x^j\\
    &= (N+1)n_x^g-\left(1+\frac{1}{N}\right)(n_x^g)^2.
    \end{split}
\end{align}

\noindent We get
\begin{align}
\begin{split}
    h_x^0&=\mu_g n_x^g+\frac{U_{gg}}{2}n_x^g(n_x^g-1)+\sum_{r}\left[\mu_e n_{x,r}^e+\frac{U_{eg}^{+} +U_{eg}^{-}}{2}n_x^g n_{x,r}^e    +\frac{U_{eg}^{+} -U_{eg}^{-}}{2}\psi_x^{i\dagger} c_{x,r}^{j \dagger}\psi_x^j c_{x, r}^{i}\right]\\
    &=\left[\mu_g-\frac{U_{gg}}{2}+\frac{U_{eg}^{+} -U_{eg}^{-}}{2}(N+1)\right] n_x^g+\frac{(N-1)U_{eg}^+ +(N+1)U_{eg}^-}{2N}\left(n_x^g+\sum_r n_{x,r}^e\right)n_x^g\\
    &\qquad+\left(\frac{U_{gg}}{2}+U_{eg}^+\right)(n_x^g)^2    -\frac{U_{eg}^{+} -U_{eg}^{-}}{4}M_x^a G_x^a.
    \end{split}
\end{align}

We 
can write $G_x=n_x^g+\sum_r n_{x,r}^e-C$ where $C$ is an appropriate constant.
With this definition, the single block Hamiltonian $h_x^0$ becomes
\begin{equation}\label{eq:demo0}
    h_x^0=\mu_g n_x^g+\mu_e(G_x-n_x^g+C)+\frac{U_{gg}}{2}n_x^g(n_x^g-1)
    +U_{eg}^+ n_x^g(G_x-n_x^g+C) -\frac{U_{eg}^+ -U_{eg}^-}{4}M^a_x G^a_x.
\end{equation}
We can then use that $\sum_x n_x^g=\text{const.}$ and $\sum_x G_x=\text{const.}$ to write
\begin{equation}
    H_0=\sum_x h_x^0=\sum_x \left[\frac{U_{gg}-2U_{eg}^+}{2}n_x^g(n_x^g-1)
    +U_{eg}^+ n_x^g G_x -\frac{U_{eg}^+ -U_{eg}^-}{4}M^a_x G^a_x\right]+\text{const.}
\end{equation}

\subsection{Example: $N=2$}
As an example, here we compute the eigenstates of $h^0_x$ on a single block for the simple case $N=2$ in one spatial dimension. For convenience of notation, we define the following quantities
\begin{equation}
    U_{eg}^n = \frac{U_{eg}^++3U_{eg}^-}{4},\qquad U_{eg}^s=\frac{U_{eg}^--U_{eg}^+}{2},
\end{equation}

\begin{equation}
    n^e_x =\sum_r n_{x,r}^e, \qquad  s_x^{g,a} =\frac{1}{2}\psi_x^{i\dagger}\sigma^a_{ij}\psi_x^j,\qquad s_x^{e,a} =\frac{1}{2}\sum_r c_{x,r}^{i\dagger}\sigma^a_{ij}c_{x,r}^j.
\end{equation}
The single block Hamiltonian can be conveniently expressed as
\begin{equation}
    h^0_x=\mu_g n^g_x+\mu_e  n_x^e+U_{gg} \frac{n^g_x(n^g_x-1)}{2}+ U_{eg}^n n^e_x n^g_x+2U_{eg}^s s_{x}^{e,a} s_x^{g,a}.
\end{equation}
In Tab. \ref{tab:h0states} we report the eigenstates of $h^0_x$, their quantum numbers, and their energies. In Appendix~\ref{app:effH} we use these states to compute the effective Hamiltonian in perturbation theory.

\begin{table}[h!]
\centering
\begin{tabular}{cccccccc}\toprule
State & $n_x^g$ & $n_x^e$ & $s_x^g$ & $s_x^e$ & $s_x^\mathrm{tot}$ & $s_{x}^{e,a} s_x^{g,a}$ & Energy \\
\toprule
 $\ket{0}$& $0$ & $0$ & $0$ & $0$ & $0$ & $0$ & 0\\
 \midrule
 $\ket{g^1},\;\ket{g^2}$ & $1$ & $0$ & $1/2$ & $0$ & $1/2$ & $0$ & $\mu^g$\\
 \midrule
 \begin{tabular}{@{}c@{}}
 $\ket{e_+^1},\;\ket{e_+^2}$,\\ $\ket{e_-^1},\;\ket{e_-^2}$
 \end{tabular}& $0$ & $1$ & $0$ & $1/2$ & $1/2$ & $0$ & $\mu^e$\\
 \midrule
 $\ket{g^1 g^2}$ & $2$ & $0$ & $0$ & $0$ & $0$ & $0$ & $2\mu^g+U_{gg}$\\
 \midrule
 \begin{tabular}{@{}c@{}}
 $\frac{1}{\sqrt{2}}(\ket{g^1 e^2_+}-\ket{g^2 e^1_+})$,\\$ \frac{1}{\sqrt{2}}(\ket{g^1 e^2_-}-\ket{g^2 e^1_-})$
 \end{tabular}& $1$ & $1$ & $1/2$ & $1/2$ & $0$ & $-3/4$ & $\mu^g+\mu^e+U_{eg}^+$\\
 \midrule
 \begin{tabular}{@{}c@{}} $\frac{1}{\sqrt{2}}(\ket{g^1 e^2_+}+\ket{g^2 e^1_+})$,\\$ \ket{g^1 e^1_+},\;  \ket{g^2 e^2_+},$\\
 $\frac{1}{\sqrt{2}}(\ket{g^1 e^2_-}+\ket{g^2 e^1_-})$,\\$ \ket{g^1 e^1_-},\;  \ket{g^2 e^2_-},$ \end{tabular} 
  & $1$ & $1$ & $1/2$ & $1/2$ & $1$ & $1/4$ & $\mu^g+\mu^e+U_{eg}^-$\\
  \midrule
 \begin{tabular}{@{}c@{}}
  $ \ket{e^1_+ e^2_+},\;  \ket{e^1_- e^2_-},$\\$\frac{1}{\sqrt{2}}(\ket{e^1_+ e^2_-}-\ket{e^1_+ e^2_-}) $
  \end{tabular}& $0$ & $2$ & $0$ & $0$ & $0$ & $0$ & $2\mu_e$\\
  \midrule
 \begin{tabular}{@{}c@{}}
   $\ket{e^1_+ e^1_-},\;  \ket{e^2_+ e^2_-},$\\$\frac{1}{\sqrt{2}}(\ket{e^1_+ e^2_-}+\ket{e^1_+ e^2_-})$
  \end{tabular}& $0$ & $2$ & $0$ & $1$ & $1$ & $0$ & $2\mu_e$\\
  \midrule
 \begin{tabular}{@{}c@{}}
  $\ket{g^1 g^2 e^1_+}, \; \ket{g^1 g^2 e^2_+},$\\
  $\ket{g^1 g^2 e^1_-}, \; \ket{g^1 g^2 e^2_+}$
  \end{tabular}& $2$ & $1$ & $0$ & $1/2$ & $1/2$ & $0$ & $2\mu^g+\mu^e+U_{gg}+2U_{eg}^n$\\
  \midrule
 \begin{tabular}{@{}c@{}}
  $\ket{g^1 e^1_+ e^2_+}, \; \ket{g^2 e^1_+ e^2_+},$\\
  $\ket{g^1 e^1_- e^2_-}, \; \ket{g^2 e^1_- e^2_-}$
  \end{tabular}& $1$ & $2$ & $0$ & $1/2$ & $1/2$ & $0$ & $\mu^g+2\mu^e+2U_{eg}^n$\\
  \midrule
 \begin{tabular}{@{}c@{}}
  $\ket{e^1_+, e^2_+, e^1_-}, \; \ket{e^1_+, e^2_+, e^2_-},$\\
  $\ket{e^1_+, e^1_-, e^2_-}, \; \ket{e^2_+, e^1_-, e^2_-}$
  \end{tabular}& $0$ & $3$ & $0$ & $1/2$ & $1/2$ & $0$ & $3\mu^e$\\
  \midrule
 \begin{tabular}{@{}c@{}}
  $\frac{1}{\sqrt{2}}(\ket{g^1 e^1_+ e^2_-}-\ket{g^1 e^2_+ e^2_-}),$\\
  $\frac{1}{\sqrt{2}}(\ket{g^2 e^1_+ e^2_-}-\ket{g^2 e^2_+ e^2_-})$
  \end{tabular}& $1$ & $2$ & $1/2$ & $0$ & $1/2$ & $0$ & $\mu^g+2\mu^e+2U_{eg}^n$\\
  \midrule
 \begin{tabular}{@{}c@{}}
  $\ket{g^1 e^1_+ e^1_-},\;\ket{g^2 e^2_+ e^2_-},$\\
  $\frac{1}{\sqrt{3}}(\ket{g^1 e^2_+ e^2_-}+\ket{g^2 e^1_+ e^2_-}+\ket{g^2 e^2_+ e^1_-}),$\\
  $\frac{1}{\sqrt{3}}(\ket{g^2 e^1_+ e^1_-}+\ket{g^1 e^2_+ e^1_-}+\ket{g^1 e^1_+ e^2_-})$
  \end{tabular}& $1$ & $2$ & $1/2$ & $1$ & $3/2$ & $1/2$ & $\mu^g+2\mu^e+2U_{eg}^-$\\
  \midrule
 \begin{tabular}{@{}c@{}}
  $\frac{1}{\sqrt{6}}(2\ket{g^1 e^2_+ e^2_-}-\ket{g^2 e^1_+ e^2_-}-\ket{g^2 e^2_+ e^1_-})$\\
  $\frac{1}{\sqrt{6}}(2\ket{g^2 e^1_+ e^1_-}-\ket{g^1 e^2_+ e^1_-}-\ket{g^1 e^1_+ e^2_-})$
  \end{tabular}& $1$ & $2$ & $1/2$ & $1$ & $1/2$ & $-1$ & $\mu^g+2\mu^e+2(U_{eg}^n-U_{eg}^s)$\\
  \bottomrule
    \end{tabular}
    \caption{Eigenstates of $h^0$ for number of particles $n_x^e+n_x^g\le 3$. The notation implies proper antisymmetrization of the fermionic wavefunction (for example $\ket{g^1g^2e_+^2}\equiv \psi_x^{1\dagger}\psi_x^{2\dagger}c_{x,+}^{2\dagger}\ket{0}$).}
    \label{tab:h0states}
\end{table}

\section{Effective Hamiltonian}
\label{app:effH}
We here derive the effective Hamiltonian to second order in perturbation theory for the case $N=2$ in one spatial dimension. The perturbation can be written as a sum of terms that act over pairs of neighboring blocks:
\begin{equation}
    H_1=\sum_x h_{x,x+1}^1,
\end{equation}
\begin{multline}
    h_{x,x+1}^1=-t_g (\psi^{i\dagger}_x \psi^i_{x+1}+\text{H.c.})-t_e (c^{i\dagger}_{x,+} c^i_{x+1,-}+\text{H.c.})
    +\frac{A_{eg}^+ +A_{eg}^-}{2}(n_{x,+}^e+n_{x+1,-}^e)(\psi_x^{i\dagger}\psi_{x+1}^i +\text{H.c.})\\
    +\frac{A_{eg}^- -A_{eg}^+}{2}(c_{x,+}^{i\dagger}c_{x,+}^j\psi_x^{j\dagger}\psi_{x+1}^i+c_{x+1,-}^{i\dagger}c_{x+1,-}^j\psi_x^{j\dagger}\psi_{x+1}^i +\text{H.c.}),
\end{multline}
where longer range terms and other interactions are neglected (we check the validity of this approximation for realistic parameters in Appendix \ref{app:abinitio}).
It is easy to check that $h_{x,x+1}$ has no matrix elements between gauge-invariant states. The second-order effective Hamiltonian on the gauge-invariant subspace can then be obtained then as a sum of two body operators
\begin{equation}
    H_\text{eff}=\sum_x h_{x,x+1}^\text{eff},\qquad \bra{j}{h_{x,x+1}^\text{eff}}\ket{l}=\sum_{k }(\bra{k}{h_{x,x+1}^1}\ket{j})^*\bra{k}{h_{x,x+1}^1}\ket{l}\omega_{jk}^{-1},
\end{equation}
where $\ket{j}$ and $\ket{l}$ belong to the Hilbert space of two blocks and can be represented as the tensor product of gauge-invariant eigenstates of $h^0$, with the additional constraint that the link connecting them hosts a single $e$ particle. The states $\ket{k}$ also belong to the Hilbert space of two blocks, but are the tensor products of eigenstates of $h^0$ that are not gauge-invariant. The states $\ket{j}$ and $ \ket{k}$ are degenerate eigenstates of $h_x^0+h_x^1$ (we assume the resonance condition $U_{gg}=2U_{eg}^+$), with eigenvalue $\epsilon_j$, while the intermediate state $\ket{k}$ has eigenvalue $\epsilon_k$. We defined $\omega_{jk}=\epsilon_j-\epsilon_k$.

Since $h_{x,x+1}^1$ conserves the total number of $g$ atoms and the total number of $e$ atoms on the two blocks, the effective Hamiltonian has a block structure, where each block is labelled by these two numbers of atoms. We now compute the effective Hamiltonian separately for each of these blocks.

\subsection{Sector with 3 $g$ atoms and 1 $e$ atom}
A basis for the gauge-invariant states on this block is defined by the two states 

\begin{equation}
    \ket{0_\text{gi}^{(3,1)}}=\frac{1}{\sqrt{2}}\epsilon_{ab} \ket{g^1g^2}_x \ket{g^a e_-^b}_{x+1},\qquad \ket{1_\text{gi}^{(3,1)}}=\frac{1}{\sqrt{2}}\epsilon_{ab}\ket{g^a e_+^b}_x \ket{g^1g^2}_{x+1},
\end{equation}
where we defined
\begin{align}
\epsilon_{ab}=\begin{cases}
1 \;& a=1,b=2\\
-1\; &a=2, b=1\\
0\; &a=b.
\end{cases}
\end{align}

The non-zero matrix elements of $h_{x,x+1}^{(1)}$ between these states and non-gauge invariant states take the form

\begin{equation}
\left(h_{x,x+1}^1\right)_{kj}^{(3,1)}=
    \begin{pmatrix}
        t_g+\dfrac{1}{2}(A_{eg}^+ + A_{eg}^-) & -t_e\\
        -t_e & t_g+\dfrac{1}{2}(A_{eg}^+ + A_{eg}^-)
    \end{pmatrix},
\end{equation}
where the column index $j=0,1$ labels the gauge-invariant basis states $\ket{j_\text{gi}^{(3,1)}}$ defined above, while the row index $k=0,1$ is for the non-gauge-invariant states $\ket{k_\text{ngi}^{(3,1)}}$
\begin{equation}
    \ket{0_\text{ngi}^{(3,1)}}=\frac{1}{\sqrt{2}}\epsilon_{ab}\ket{g^a}_x\ket{g^1 g^2e^b_-}_{x+1},\qquad \ket{1_\text{ngi}^{(3,1)}}=\frac{1}{\sqrt{2}}\epsilon_{ab}\ket{e^a_+}_x\ket{g^1 g^2 e^b_-}_{x+1}.
\end{equation}
 
For both values of $k$, one finds $\omega_{j,k}=U_{eg}^+-2U_{eg}^n=(U_{eg}^+-3U_{eg}^{-})/2$. The effective Hamiltonian in this sector then reads

\begin{equation}
\label{eq:heff31}
    \left(h_{x,x+1}^\text{eff}\right)^{(3,1)}_{j_1,j_2}=\dfrac{2}{U_{eg}^+-3U_{eg}^-}\begin{pmatrix}
    \left(t_g+\frac{A_{eg}^++A_{eg}^-}{2}\right)^2+t_e^2 & -t_e (2t_g+A_{eg}^+ +A_{eg}^-)\\
    -t_e (2t_g+A_{eg}^+ +A_{eg}^-) & \left(t_g+\frac{A_{eg}^++A_{eg}^-}{2}\right)^2+t_e^2
\end{pmatrix}
\end{equation}

\subsection{Sector with 2 $g$ atoms and 2 $e$ atoms}
Similarly to the previous case, we define a basis $\ket{j_\text{gi}^{(2,2)}}$ for the gauge-invariant states 

\begin{equation}
    \ket{0_\text{gi}^{(2,2)}}=\frac{1}{\sqrt{2}} \epsilon_{ab}\ket{g^1g^2}_x \ket{e_+^a e_-^b}_{x+1},\qquad
    \ket{1_\text{gi}^{(2,2)}}=\frac{1}{2}\epsilon_{ab}\epsilon_{cd}\ket{g^a e_+^b}_{x} \ket{g^c e_+^d}_{x+1},
\end{equation}
and a basis $\ket{k_\text{ngi}^{(2,2)}}$ for the non-gauge invariant states: 

\begin{equation}
    \ket{0_\text{ngi}^{(2,2)}}=\frac{1}{2}\epsilon_{ab}\epsilon_{cd}\ket{g^a}_x\ket{g^b e^c_+ e^d_-}_{x+1},\qquad
    \ket{1_\text{ngi}^{(2,2)}}=\frac{1}{\sqrt{2}}\epsilon_{ab}\ket{g^1g^2e^a_+}_x\ket{e^b_+}_{x+1},\nonumber
\end{equation}
\begin{equation}
    \ket{2_\text{ngi}^{(2,2)}}=\frac{1}{2\sqrt{3}}(1-\delta_{ab})\ket{g^a}_x\left(2\ket{g^a e^b_+ e^b_-}_{x+1}-\ket{g^b e^a_+ e^b_-}_{x+1}-\ket{g^b e^b_+ e^a_-}_{x+1}\right),\nonumber
\end{equation}
\begin{equation}
    \ket{3_\text{ngi}^{(2,2)}}=\frac{1}{\sqrt{2}}\epsilon_{ab}\ket{e^a_+}_x\ket{g^1 g^2 e^b_-}_{x+1}.
\end{equation}

Since $h_{x,x+1}^1$ does not change the occupations $n_{x,-}^e$ and $n_{x+1, +}^e$, so we only consider here the two states that have $n_{x,-}^e=0, n_{x+1, +}^e=1$ without loss of generality. Because of inversion symmetry, the results can be directly generalised to the sector with $n_{x,-}^e=1, n_{x+1, +}^e=0$.
The matrix elements of $h_{x,x+1}^1$ in this basis read

\begin{equation}
    \left(h_{x,x+1}^1\right)_{kj}^{(2,2)}=\begin{pmatrix}
        -\dfrac{1}{2\sqrt{2}}(4t_g +3A_{eg}^- +A_{eg}^+) & \dfrac{1}{2}t_e\\
        -t_e & \dfrac{1}{2\sqrt{2}}(2t_g + A_{eg}^- +A_{eg}^+)\\
        \dfrac{1}{2\sqrt{6}} (A_{eg}^+ -A_{eg}^-) & -\dfrac{\sqrt{3}}{2} t_e\\
        0 & \dfrac{1}{\sqrt{2}}(t_g+A_{eg}^-)
    \end{pmatrix},
\end{equation}
and the energy differences are
\begin{align}
\omega_{j,0}&=2U_{eg}^+-2U_{eg}^n=3(U_{eg}^+-U_{eg}^-)/2,\\
\omega_{j,1}&=2U_{eg}^+-U_{gg}-2U_{eg}^n=-(U_{eg}^++3U_{eg}^-)/2, \\
\omega_{j,2}&=2U_{eg}^+-2U_{eg}^n+2U_{eg}^s=(U_{eg}^+-U_{eg}^-)/2,\\
\omega_{j,3}&=2U_{eg}^+-U_{gg}-2U_{eg}^n=-(U_{eg}^++3U_{eg}^-)/2.
\end{align}
We then get the effective Hamiltonian for this sector
\begin{equation}
\label{eq:heff22}
 \left(h_{x,x+1}^\text{eff}\right)^{(2,2)}_{j_1,j_2}
    =\begin{pmatrix}
    \frac{(A_{eg}^+-A_{eg}^-)^2+(4t_g+3A_{eg}^-+A_{eg}^+)^2}{12(U_{eg}^+-U_{eg}^-)} -\frac{2t_e^2}{(U_{eg}^+ +3U_{eg}^-)} & -\frac{\sqrt{2} t_e (t_g+A_{eg}^+)}{3(U^+_{eg}-U^-_{eg})}+\frac{t_e(A_{eg}^++A_{eg}^-+2t_g)}{\sqrt{2}(U^+_{eg}+3 U^-_{eg})}\\
    -\frac{\sqrt{2} t_e (t_g+A_{eg}^+)}{3(U^+_{eg}-U^-_{eg})}+\frac{t_e(A_{eg}^++A_{eg}^-+2t_g)}{\sqrt{2}(U^+_{eg}+3 U^-_{eg})} & -\frac{4(t_g+A_{eg}^-)^2+(2t_g+A_{eg}^-+A_{eg}^+)^2}{4(U^+_{eg} + 3U^-_{eg})}+\frac{5 t_e^2}{3(U^+_{eg}-U^-_{eg})}.
\end{pmatrix}
\end{equation}

\subsection{Sector with 1 $g$ atom and 3 $e$ atoms}
We now apply the same procedure to the sector with 1 $g$ atom and 3 $e$ atoms. 
The basis states $\ket{j_\text{gi}^{(1,3)}}$ for the gauge-invariant subspace in this sector are
\begin{equation}
   \ket{0_\text{gi}^{(1,3)}}= \frac{1}{2}\epsilon_{ab}\epsilon_{cd}\ket{g^a e_-^b}_{x} \ket{e_+^c e_-^d}_{x+1} ,\qquad
   \ket{1_\text{gi}^{(1,3)}}=\frac{1}{2}\epsilon_{ab}\epsilon_{cd}\ket{e_+^a e_-^b}_{x} \ket{g^c e_+^d}_{x+1}.
\end{equation}
We then define the non-gauge-invariant states $\ket{k_\text{ngi}^{(1,3)}}$ as

\begin{equation}
    \ket{0_\text{ngi}^{(1,3)}}=\frac{1}{2}\epsilon_{ab}\epsilon_{cd}\ket{g^a e^c_+ e^d_-}_x\ket{e^b_+}_{x+1},\qquad
    \ket{1_\text{ngi}^{(1,3)}}=\frac{1}{2}\epsilon_{ab}\epsilon_{cd}\ket{e^a_-}_x\ket{g^b e^c_+ e^d_-}_{x+1},\nonumber
\end{equation}
\begin{equation}
    \ket{2_\text{ngi}^{(1,3)}}=\frac{1}{2\sqrt{3}}(1-\delta_{ab})\left(2\ket{g^a e^b_+ e^b_-}_x-\ket{g^b e^a_+ e^b_-}_{x+1}-\ket{g^b e^b_+ e^a_-}_{x+1}\right)\ket{e^a_+ }_{x+1},\nonumber
\end{equation}
\begin{equation}
    \ket{3_\text{ngi}^{(1,3)}}=\frac{1}{2\sqrt{3}}(1-\delta_{ab})\ket{e^a_-}_x\left(2\ket{g^a e^b_+ e^b_-}_{x+1}-\ket{g^b e^a_+ e^b_-}_{x+1}-\ket{g^b e^b_+ e^a_-}_{x+1}\right).
\end{equation}
We compute the matrix elements of $h_{x,x+1}^1$:
\begin{equation}
    \left(h_{x,x+1}^1\right)_{kj}^{(1,3)}=\begin{pmatrix}
        \dfrac{t_e}{2} & -t_g-\dfrac{1}{4}(A_{eg}^+ +3 A_{eg}^-)\\
        -t_g-\dfrac{1}{4}(A_{eg}^+ +3 A_{eg}^-) & \dfrac{t_e}{2}\\
        -\dfrac{\sqrt{3}}{2} t_e & \dfrac{1}{4\sqrt{3}} (A_{eg}^+-A_{eg}^-)\\
        \dfrac{1}{4\sqrt{3}} (A_{eg}^+-A_{eg}^-) & -\dfrac{\sqrt{3}}{2} t_e
    \end{pmatrix},
\end{equation}
and the energy differences
\begin{align}
\omega_{j,0}=\omega_{j,1}&=U_{eg}^+-2U_{eg}^n=(U_{eg}^+-3U_{eg}^-)/2,\\
\omega_{j,2}=\omega_{j,3}&=U_{eg}^+-2U_{eg}^n+2U_{eg}^s=-(U_{eg}^++U_{eg}^-)/2.
\end{align}
We finally get the effective Hamiltonian in this sector

\begin{equation}
\label{eq:heff13}
\left(h_{x,x+1}^\text{eff}\right)^{(1,3)}_{j_1,j_2}=\begin{pmatrix}
    \frac{(4t_g+A_{eg}^++3A_{eg}^-)^2 +4t_e^2}{8(U^+_{eg}-3U^-_{eg})} -\frac{36 t_e^2+(A_{eg}^+-A_{eg}^-)^2}{24(U^+_{eg}+U^-_{eg})} & \frac{t_e(A_{eg}^+-A_{eg}^-)}{2(U_{eg}^+ + U_{eg}^-)}
    -\frac{t_e(4t_g+A_{eg}^++3A_{eg}^-)}{2(U^+_{eg}-3U^-_{eg})}\\
    \frac{t_e(A_{eg}^+-A_{eg}^-)}{2(U_{eg}^+ + U_{eg}^-)}
    -\frac{t_e(4t_g+A_{eg}^++3A_{eg}^-)}{2(U^+_{eg}-3U^-_{eg})} & \frac{(4t_g+A_{eg}^++3A_{eg}^-)^2 +4t_e^2}{8(U^+_{eg}-3U^-_{eg})} -\frac{36 t_e^2+(A_{eg}^+-A_{eg}^-)^2}{24(U^+_{eg}+U^-_{eg})}  .
\end{pmatrix}
\end{equation}

\subsection{Parameters of the effective Hamiltonian}
Once we have the effective 2-body Hamiltonian $h_{x,x+1}$ in its matrix forms, with the blocks defined in Eqs.~(\ref{eq:heff31}), (\ref{eq:heff22}) and (\ref{eq:heff13}), it is useful to express it in terms of the local densities and creation/annihilation operators. To this end, we split $h_{x,x+1}$ in its diagonal and off-diagonal parts 
\begin{equation}
    h_{x, x+1}^\text{eff}=h_{x, x+1}^\text{eff, diag}+h_{x, x+1}^\text{eff, off-diag},
\end{equation}
and we make the following ansatz for the diagonal part:
\begin{equation}
\label{eq:heffdiag}
    h_{x, x+1}^\text{eff,diag}=w n_x^g n_{x+1}^g+\frac{1}{2}\sum_{y=x,x+1}\left(\tilde{\mu}_g n_y^g+\tilde{\mu}_e n_{y,+}^e+\tilde{\mu}_e n_{y,-}^e+\frac{u}{2}n_y^g(n_y^g-1)\right).
\end{equation}

The coefficients $\tilde \mu_g, \tilde \mu_e, u, w$ can then be obtained by evaluating Eq.~\eqref{eq:heffdiag} for the gauge-invariant basis states defined in each sector, and setting them to be equal to the diagonal matrix elements computed above. We thus have to solve the linear system of equations
\begin{equation}
\begin{dcases}
    \left(h_{x, x+1}^\text{eff}\right)^{(3,1)}_{0,0}=
    2w+\frac{3}{2}\tilde \mu_g +\frac{1}{2}\tilde \mu_e+\frac{u}{2}\\
    \left(h_{x, x+1}^\text{eff}\right)^{(2,2)}_{0,0}=
    \tilde \mu_g +\tilde \mu_e+\frac{u}{2}\\
    \left(h_{x, x+1}^\text{eff}\right)^{(2,2)}_{1,1}=
    w+\tilde \mu_g +\tilde \mu_e\\
    \left(h_{x, x+1}^\text{eff}\right)^{(1,3)}_{0,0}=
    \frac{1}{2}\tilde \mu_g +\frac{3}{2}\tilde \mu_e.
    \end{dcases}
\end{equation}
Since the total numbers of $g$ and $e$ atoms are conserved, the terms $\tilde \mu_g$ and $\tilde \mu_g$ only contribute as inconsequential constants to the effective Hamiltonian. The on-site and nearest-neighbor mater interactions ($u$ and $w$ respectively) play a role in the dynamics. The solution of the linear systems yields

\begin{equation}
u=     2\left(h_{x, x+1}^\text{eff}\right)^{(3,1)}_{0,0}-4\left(h_{x, x+1}^\text{eff}\right)^{(2,2)}_{1,1}+2\left(h_{x, x+1}^\text{eff}\right)^{(1,3)}_{0,0},
\end{equation}
\begin{equation}
w=  \left(h_{x, x+1}^\text{eff}\right)^{(3,1)}_{0,0}-\left(h_{x, x+1}^\text{eff}\right)^{(2,2)}_{0,0}-\left(h_{x, x+1}^\text{eff}\right)^{(2,2)}_{1,1}+\left(h_{x, x+1}^\text{eff}\right)^{(1,3)}_{0,0}.
\end{equation}
Similarly to the diagonal part, we formulate the following ansatz for the off-diagonal terms:
\begin{equation}
    h_{x, x+1}^\text{eff, off-diag}= -[\tau_1 +\tau_2 E_{x-1,x} E_{x+1,x+2}+\tau_3 (E_{x-1,x}- E_{x+1,x+2})](\psi_x^{i\dagger} U_{x,x+1}^{ij} \psi_{x+1}^j +\text{H.c.}).
\end{equation}
The system of equations that we aim to solve is
\begin{equation}
    \begin{dcases}
        \left(h_{x, x+1}^\text{eff}\right)^{(3,1)}_{0,1}=-2\left(\tau_1-\frac{1}{4}\tau_2-\tau_3\right)\\
        \left(h_{x, x+1}^\text{eff}\right)^{(2,2)}_{0,1}=-\sqrt{2}\left(\tau_1+\frac{1}{4}\tau_2\right)\\
        \left(h_{x, x+1}^\text{eff}\right)^{(1,3)}_{0,1}=-\left(\tau_1-\frac{1}{4}\tau_2+\tau_3\right).
    \end{dcases}
\end{equation}
We obtain
\begin{align}
    \tau_1&=-\frac{1}{8}\left(h_{x, x+1}^\text{eff}\right)^{(3,1)}_{0,1}-\frac{1}{2\sqrt{2}}\left(h_{x, x+1}^\text{eff}\right)^{(2,2)}_{0,1}-\frac{1}{4}\left(h_{x, x+1}^\text{eff}\right)^{(1,3)}_{0,1},\\
    \tau_2&=\frac{1}{2}\left(h_{x, x+1}^\text{eff}\right)^{(3,1)}_{0,1}-\sqrt{2}\left(h_{x, x+1}^\text{eff}\right)^{(2,2)}_{0,1}+\left(h_{x, x+1}^\text{eff}\right)^{(1,3)}_{0,1},\\
    \tau_3&=\frac{1}{4}\left(h_{x, x+1}^\text{eff}\right)^{(3,1)}_{0,1}-\frac{1}{2}\left(h_{x, x+1}^\text{eff}\right)^{(1,3)}_{0,1}.
\end{align}

\section{Initial state preparation}
\label{sec:stateprep}

\begin{figure}[htb]
    \centering
    \includegraphics[width=0.9\linewidth]{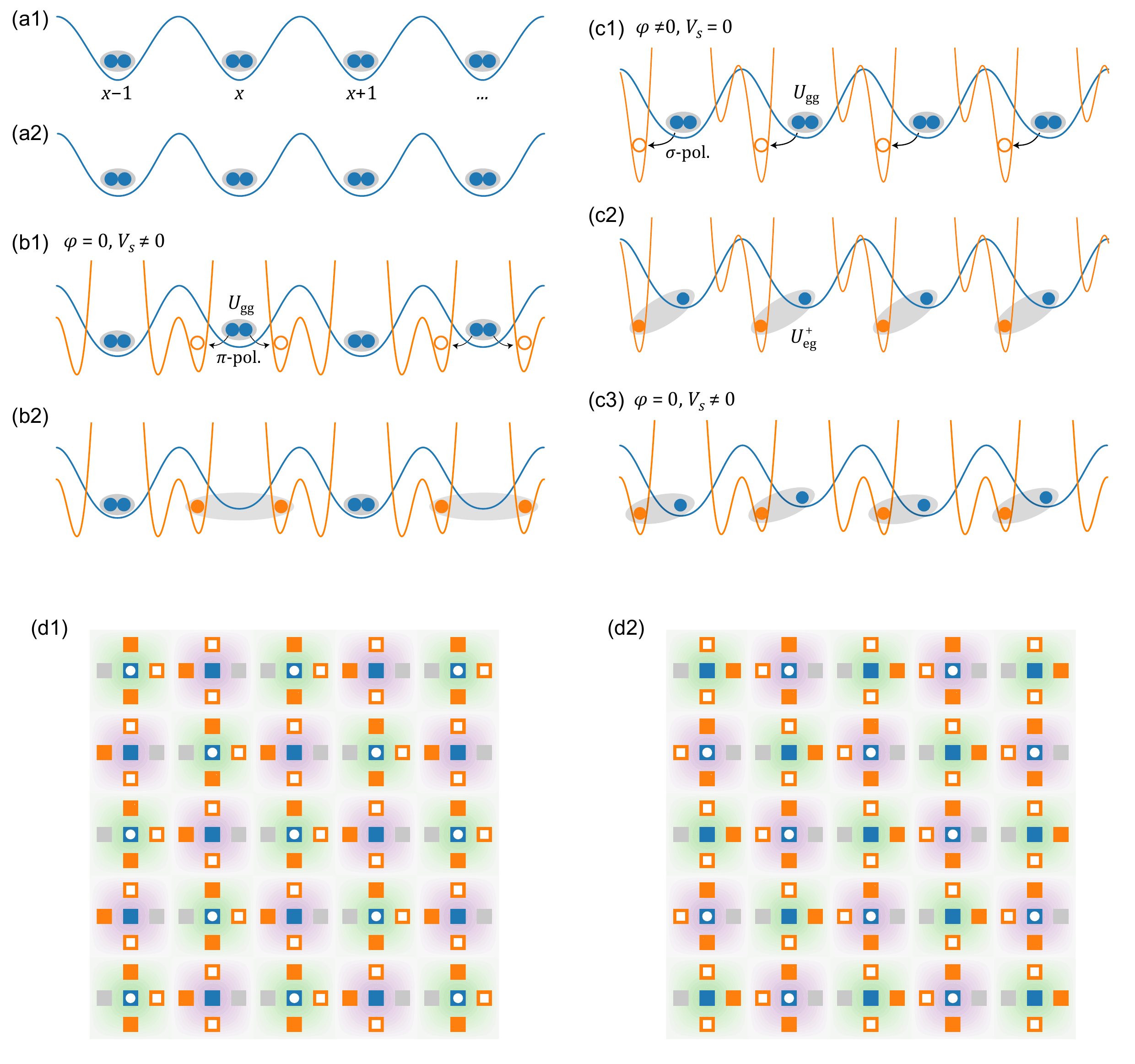}
    \caption{Experimental protocols for gauge-invariant state preparation. (a1-a2) A $SU(N=2)$ band insulator of $g$-state atoms is prepared first in the long lattice $V^0_\alpha$, and the short lattice $V^1_\alpha$ is subsequently added. (b1-b2) Preparation sequence for obtaining a gauge-invariant staggered block configuration. A finite shift $\delta_e - \delta_g \neq 0$ created by the staggering lattice $V^S_\alpha$ allows for selecting only even or odd blocks, where the onsite $\ket{g,g}$ singlet is transferred to $\left(\ket{e+,e-}+\ket{e-, e+}\right)$, occupying both left and right rishon sites of the block. (c) Preparation sequence for obtaining a homogeneous gauge-invariant configuration. A relative phase $\varphi \neq 0$ between the long $V^0_\alpha(x)$ and the short $V^1_\alpha(x)$ lattices allows for transferring on each block the onsite $\ket{g,g}$ singlet to a $e-g$ singlet, occupying the left rishon site and possessing an interaction energy $U_{eg}^+$. Subsequently, $\varphi$ is ramped to 0, and the staggering lattice $V^S_\alpha$ is turned on adiabatically.
    (d) The combination of two 1D protocols illustrated above allows to initialize gauge-invariant states in a 2D brick-wall lattice (greyed-out orange boxes highlight double wells that are initialized with no $e$ atoms).}
    \label{fig:prep-protocols}
\end{figure}

In order to initialize the atomic system such that only gauge-invariant (spin singlet) states are occupied, it is convenient to prepare first a one-dimensional $g$-state band-insulating phase in the $a_0$-spacing lattice $V_g^0(x)$. Thereby, each site of the lattice is occupied by $N$ atoms in a spin antisymmetric state. For $N=2$, this corresponds to the $\ket{g,g}$ state. Let us now consider the available two-particle spin-singlet states for a certain block in 1D, i.e.:
\begin{equation}
\left\{ \ket{g,g}, \frac{1}{\sqrt{2}}\left(\ket{g,e\pm}+\ket{e\pm, g}\right), \frac{1}{\sqrt{2}}\left(\ket{e+,e-}+\ket{e-, e+}\right) \right\}.
\end{equation}
Among these, states with different electronic state populations can be coupled together by laser light driving the ${}^1$S$_0 \rightarrow {}^3$P$_0$ clock transition, suitably polarized and tuned to be selectively resonant with the transitions between different two-particle states \cite{Scazza14}, which are split owing to the different interaction strenghts in the singlet and triplet channels. In particular, the relevant Rabi couplings for $\pi$-polarized clock light are given by:
\begin{subequations}
\begin{align}
\frac{\hbar}{2}\Omega^{eg\pm} &= \left(\bra{g,e\pm}\pm\bra{e\pm, g}\right) H_\pi^{(2)} \ket{g,g}= \frac{\hbar}{\sqrt{2}}(\Omega_\uparrow \mp\Omega_\downarrow)\\[0.5mm]
\frac{\hbar}{2}\Omega^{ee} &= \left(\bra{g,e\pm}\pm\bra{e\pm, g}\right) H_\pi^{(2)} \left(\ket{e+,e-}+\ket{e-, e+}\right) = - \frac{\hbar}{\sqrt{2}}(\Omega_\uparrow \mp\Omega_\downarrow)
\end{align}
\end{subequations}
where $H_\pi^{(2)}$ is the two-particle atom-light interaction Hamiltonian with clock $\pi$-polarized light, 
and $\Omega_i$ is the single-particle clock Rabi frequency of spin state $i=\uparrow,\downarrow$ with $\pi$ polarization. 
It is thus possible to selectively couple all onsite $\ket{g,g}$ pairs to the desired singlet state by means of a global clock light pulse. The relative phase $\varphi$ between the long and the short lattice potentials $V_\alpha^0(x)$ and $V_\alpha^1(x)$ can be further exploited to selectively prepare the states $\left(\bra{g,e\pm}+\bra{e\pm, g}\right)/\sqrt{2}$. The experimental protocols to create the $\left(\ket{e+,e-}+\ket{e-, e+}\right)/\sqrt{2}$ on each odd or even block (see Fig.~1(c) of the main text), or the $\left(\bra{g,e-}+\bra{e-, g}\right)$ state on each block are sketched in Fig.~\ref{fig:prep-protocols}.

\end{document}